\documentclass[manuscript=article,layout=twocolumn]{achemso}

\usepackage{graphicx}
\usepackage{dcolumn}
\usepackage{bm}
\usepackage[utf8]{inputenc}
\DeclareUnicodeCharacter{2010}{-}
\usepackage[T1]{fontenc}
\usepackage{mathptmx}
\usepackage{booktabs}
\usepackage{color}
\usepackage{multirow}
\usepackage{notes2bib}
\providecommand{\tabularnewline}{\\}
\usepackage{textcomp}
\usepackage{physics}
\usepackage{subcaption}
\usepackage[version=4]{mhchem}
\newcommand{\angstrom}{\text{\normalfont\AA}}

\SectionNumbersOn 

\author{Juan Felipe Huan Lew-Yee}
\affiliation{Departamento de F\'isica y Qu\'imica Te\'orica, Facultad de Qu\'imica,
Universidad Nacional Aut\'onoma de M\'exico, M\'exico City, C.P. 04510,
M\'exico}

\author{Jorge M. del Campo}
\email{jmdelc@unam.mx}
\affiliation{Departamento de F\'isica y Qu\'imica Te\'orica, Facultad de Qu\'imica,
Universidad Nacional Aut\'onoma de M\'exico, M\'exico City, C.P. 04510,
M\'exico}

\author{Mario Piris}
\email{mario.piris@ehu.eus}
\affiliation{Kimika Fakultatea, Euskal Herriko Unibertsitatea (UPV/EHU),
P.K. 1072, 20080 Donostia, Euskadi (Spain)}
\alsoaffiliation{Donostia International Physics Center (DIPC), 20018 Donostia, Euskadi (Spain).}
\alsoaffiliation{IKERBASQUE, Basque Foundation for Science, 48013 Bilbao, Euskadi, (Spain).}

\title{Electron correlation in the Iron(II) Porphyrin by NOF approximations}

\begin{document}

\begin{tocentry}

\includegraphics[]{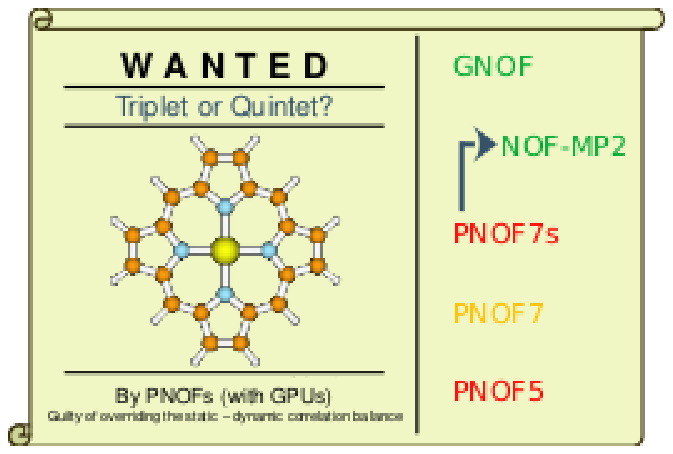}

\end{tocentry}

\begin{abstract}

The relative stability of the singlet, triplet, and quintet spin states of Iron(II) porphyrin (FeP) represents a challenging problem for electronic structure methods. While it is currently accepted that the ground state is a triplet, multiconfigurational wavefunction-based methods predict a quintet, and density functional approximations vary between triplet and quintet states, leading to a prediction that highly depends on the features of the method employed. The recently proposed Global Natural Orbital Functional (GNOF) aims to provide a balanced treatment between static and dynamic correlation, and together with the previous Piris Natural Orbital Functionals (PNOFs), allowed us to explore the importance of each type of correlation in the stability order of the states of FeP with a method that conserves the spin of the system. It is noteworthy that GNOF correlates all electrons in all available orbitals for a given basis set; in the case of the FeP with a double zeta basis set as used in this work; this means that GNOF can properly correlate 186 electrons in 465 orbitals, significantly increasing the sizes of systems amenable to multiconfigurational treatment. Results show that PNOF5, PNOF7s and PNOF7 predict the quintet to have a lower energy than the triplet state; however, the addition of dynamic correlation via second-order M\o ller\textendash Plesset corrections (NOF-MP2) turns the triplet state to be lower than the quintet state, a prediction also reproduced by GNOF that incorporates much more dynamic correlation than its predecessors.

\end{abstract}

\section{Introduction}

As early as the 1970s, it was suggested that one-particle reduced density matrix (RDM) functional theory\citep{Gilbert1975,Donnelly1978,Levy1979,Valone1980} could be an attractive alternative formalism to wavefunction-based methods. Unfortunately, calculations based on exact functionals generated by the constrained-search formulation are computationally too expensive, which has prompted the development of approximate functionals for practical applications. The functionals currently in use are constructed on the basis where the one-particle RDM is diagonal, which is the definition of a natural orbital functional (NOF).\citep{Goedecker2000,Piris2007} In fact, it is more appropriate to speak of a NOF rather than a one-particle RDM functional when dealing with approximate functionals, since a two-particle RDM dependence persists\citep{Donnelly1979} and leads to the functional N-representability problem.\citep{Ludena2013,Piris2018d} An extensive account on the evolution of approximate NOFs up to the year 2018 can be found elsewhere.\citep{Piris2014a,Pernal2016,Schade2017,MITXELENA2019} 

Recent developments\citep{Piris2019,Benavides-Riveros2019,Cioslowski2019,Giesbertz2019,Gritsenko2019,Lopez2019,Quintero-Monsebaiz2019,Schilling2019,Schmidt2019,buchholz2019,Benavides-Riveros2020,Giesbertz2020,Cioslowski2020,Mitxelena2020,Mitxelena2020b,Mitxelena2020c,Lew-Yee2021,Mercero2021,Quintero-Monsebaiz2021,Schilling2021,Liebert2021,Wang2021,Yao2021,Gibney2021,Piris2021b,DiSabatino2022,Lemke2022,Mitxelena2022,Liebert2022,Wang2022,Ding2022,Rodriguez-Mayorga2022,Senjean2022,Lew-Yee2022} show that NOF theory has become an active field of research. Nowadays, an open-source implementation of NOF-based methods is available (github.com/DoNOF) to the scientific community. The associated computer program DoNOF (Donostia Natural Orbital Functional)\citep{Piris2021} is designed to solve the energy minimization problem of an approximate NOF, describing the ground state of an N-electron system in terms of natural orbitals (NOs) and their occupation numbers (ONs). Fractional occupancies naturally allow NOFs to recover the static correlation. In fact, approximate NOFs have demonstrated \citep{Mitxelena2017a,Mitxelena2022} to be more accurate than their electron density-dependent counterparts for highly multiconfigurational systems and scale satisfactorily compared to wavefunction-type methods with respect to the number of basis functions.

Particularly successful in describing static electronic correlation are electron-pairing-based NOFs,\citep{Piris2018e} namely PNOF5,\citep{Piris2011,Piris2013e} PNOF6,\citep{Piris2014c} and PNOF7.\citep{Piris2017,mitxelena2018a} For instance, PNOF6 dissociation curve of the carbon dimer closely resembles that obtained from the optimized complete active space self-consistent field wavefunction.\citep{Piris2016} So far, only NOFs that satisfy the electron-pairing constraints have provided the correct number of electrons in the fragments after homolytic dissociation.\citep{Matxain2011,Ruiperez2013} PNOF5 - PNOF7 take into account most of the non-dynamical effects, and also an important part of the dynamic electron correlation corresponding to the intrapair interactions, hence they produce results that are in good agreement with accurate wavefunction-based methods for small systems, where electron correlation effects are almost entirely intrapair. However, when the number of pairs increases, the total energy values deteriorate, especially in those regions where dynamic correlation prevails.

There are several strategies for adding the missing dynamic correlation to an approximate NOF, but second-order perturbative corrections are probably the simplest and cheapest way to properly incorporate dynamical correlation effects, which has given rise to two methods. The first uses a size-consistent multiconfigurational second-order perturbation theory (PT2), taking as reference the generating wavefunction of PNOF5, which leads to the PNOF5-PT2 method.\citep{Piris2013c,Piris2014b} The other proposal, called NOF-MP2,\citep{Piris2017} adds second-order M\o ller\textendash Plesset (MP2) corrections to a reference Slater determinant wavefunction formed with the NOs of PNOF7. Let us note that PNOF5 is strictly N-representable, i.e. the functional can be derived from a wavefunction that is antisymmetric in N-particles, so PNOF5-PT2 is well defined and the perturbative corrections are added to PNOF5 energy. On the contrary, for PNOF7 the generating wavefunction is unknown and in the NOF-MP2 method static and dynamic corrections are added to a Hartree-Fock (HF) type energy.

The reformulation\citep{Piris2018b} of NOF-MP2 based on the static part of PNOF7 (PNOF7s) and the orbital-invariant MP2 allowed us to prevent reference ONs and NOs from being spuriously influenced by non-dynamic correlation in dynamic correlation domains, and extend the NOF-MP2 method to any type of orbitals, including localized ones, respectively. NOF-MP2 has been shown to provide quantitative agreement for dissociation energies, with performance comparable to that of the accurate complete active space second-order perturbation theory in hydrogen abstraction reactions,\citep{Lopez2019} and is highly reliable for accurate chemical reaction mechanistic studies in elementary reactions of transition metal compounds.\citep{Mercero2021}

A canonicalization procedure applied to the NOs gave us the possibility to combine any many-body perturbation method,\citep{Rodriguez-Mayorga2021} like random-phase approximation or coupled-cluster singles and doubles, with a NOF. The inclusion of perturbative corrections improves the absolute energies over the reference NOF values and approaches the energies obtained by accurate wavefunction-based methods; however, it does not improve the quality of the reference NOs and ONs. A full optimization would be the only way to obtain completely correlated ONs and NOs. Unfortunately, such a self-consistent procedure makes perturbative methods incredibly computationally expensive, so it is preferable to recover the missing dynamic correlation using a more general NOF than PNOF7.

An important recent development that reinforced this strategy was the implementation of the resolution of the identity approximation (RI) in DoNOF\citep{Lew-Yee2021} and in the FermiONs++ program package.\citep{Lemke2022} The RI implementation substantially reduces memory and arithmetic scaling factors in NOF calculations. Such developments have made it possible to perform calculations on large systems of chemical interest with tens of atoms, hundreds of electrons, and thousands of basis functions, for example, the 117-atom 2\textasciiacute -carbamate taxol and the 168-atom valinomycin molecule.\citep{Lemke2022}

Recently,\citep{Piris2021b} a NOF was proposed for electronic systems with any spin value regardless of the external potential, that is, a global NOF (GNOF). The adjective \textquotedblleft global\textquotedblright{} is used instead of \textquotedblleft universal\textquotedblright{} to differentiate this approximate multipurpose NOF from Valone's exact one.\citep{Valone1980} GNOF is able to achieve a balanced treatment of static and dynamic electron correlation even for those systems with significant multiconfigurational character, preserving the total spin of multiplets.\citep{Piris2019} It should be noted that the agreement obtained by GNOF with accurate wavefunction-based methods is not only for relative energies but also for absolute energies, a sign of good results for good reasons. An example is the agreement obtained between GNOF and Full Configuration Interaction (FCI) for challenging dissociation processes in one, two and three dimensions.\citep{Mitxelena2022} Nevertheless, we must point out that GNOF, like its predecessors, is not variational since only some necessary N-representability conditions have been imposed, with the sole exception of PNOF5 for which we know the generating wavefunction.

\begin{figure}[ht]
\begin{centering}
{\includegraphics[scale=0.4]{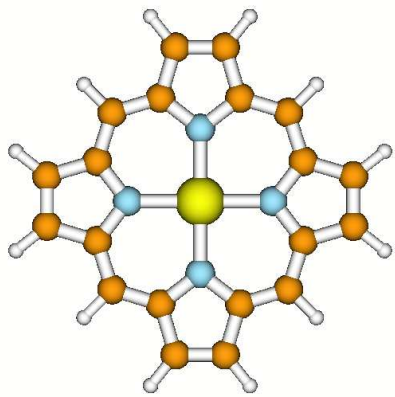}}
\caption{\label{FeP} Iron(II) Porphyrin \bigskip}
\end{centering}
\end{figure}

The simple construction of GNOF allowed us to examine the effects of different types of electron correlation. The functional has a term that fully recovers the intrapair electron correlation, that corresponds to the independent-pair model, followed by a second term that corresponds to the static interpair correlation, and it also takes into account the dynamic correlation between electron pairs. The aim of this work is to analyze the influence of different types of correlation on the spin state stability of iron(II) porphyrin molecule (FeP), as shown in Figure~\ref{FeP}, a system with 37 atoms and 186 electrons. FeP is a model system for more general substituted iron porphyrins, that play a vital role in many biological processes, including oxygen transport, electron transfer, and catalyzing the incorporation of oxygen into other molecules.\citep{Lever1989} The relationship between spin state and structure of FeP constitutes an active research topic due to its implications for the biological activity of heme proteins.\citep{Perutz1998} 

Porphyrins are molecules of chemical\citep{Gouterman1961-kl,Paolesse2017-un} and biological interest.\citep{Dayan2011-zm} However, the iron porphyrins have proven to be challenging for any theoretical method and an attractive system for testing the GNOF functional. Initial single reference studies considered a triplet state,\citep{Obara1982-kg,Sontum1983-bk,Rohmer1985-uy,Rawlings1985-sk} but subsequent multireference studies favored a quintet state.\citep{Choe1998,Choe1999} The controversy of the spin of the ground state of FeP continue up to nowadays, and the discussion has become enriched with the increase of the computational capabilities and the development of innovative methods to include more accurately the electronic correlation.

Calculations with currently used single reference methods such as coupled cluster and modern density functional approximations tends to favor the triplet as the ground state.\citep{Liao2002-zy,Groenhof2005-jp,Radon2014-th} Moreover, it has been reported that the triplet and quintet states do not present essential symmetry breaking.\citep{Lee2020-vg} In addition, typical complete active space (CAS) calculations points to the quintet,\citep{Li_Manni2016-jd} but it has been reported that increasing the size of the active space even more change the prediction to the triplet, and that the preference for a quintet may be an artifact caused by an insufficiently large active space.\citep{Smith2017a,Pierloot2017} In addition, stochastic generalized active space SCF points to the triplet.\citep{Weser2022-vl} Calculations of the recently reported pair density functional theory (PDFT) also point to the triplet.\citep{Zhou2019-rx} However, the discussion is not so easy to conclude, as density matrix renormalization group calculations (DMRG) points to a quintet state,\citep{Antalik2020-br} even after coupling with the adiabatic connection to include dynamic correlation.\citep{Beran2021-fe} Studies on the influence of the exact exchange concluded that the inclusion of large amount of it favors high-spin states, while smaller contributions favor low-spin states,\citep{Berryman2015-zv,Swart2004-aj} this becoming relevant as a recent study of PDFT has shown that the use of hybrid functionals revert the tendency to the quintet state for some on-top functional.\citep{Stroscio2022-wu} Hence, the controversy remains of active interest.

This study provides important information in many ways. First, the analysis of FeP from the perspective of PNOFs functionals might provide information on the static and dynamic correlation effects on the problem. At the same time, it will allow us to compare the set of PNOFs with the different methods previously used to study FeP. Note further that GNOF correlates all electrons into all available orbitals for a given basis set, which in the case of FeP using a double zeta basis set correlates 186 electrons in 465 orbitals. To the best of our knowledge, such a correlation calculation is not possible with current wavefunction-based methods, such as CAS or DMRG.

The work is organized as follows. First, Section~\ref{sec:theory} presents a brief review of GNOF and the M diagnostic used to characterize the NOFs solutions. This is followed by the computational details related to the NOFs calculations in Section~\ref{sec:comp-details}. Section~\ref{sec:results} presents an analysis of the performance of PNOF5, PNOF7, PNOF7s, NOF-MP2 and GNOF over the spin-stability order of FeP, together with a discussion of the electron correlation effects provided by each functional. Finally, conclusions are given in Section~\ref{sec:conclusions}.

\section{Theory}\label{sec:theory}

In this section, we briefly describe GNOF, a more detailed description can be found in Ref.~\citep{Piris2021b}. The nonrelativistic Hamiltonian under consideration is spin coordinate free; therefore, a state with total spin $S$ is a multiplet, i.e., a mixed quantum state that allows all possible $S_{z}$ values. We consider $\mathrm{N_{I}}$ single electrons which determine the spin $S$ of the system, and the rest of electrons ($\mathrm{N_{II}}=\mathrm{N-N_{I}}$) are spin-paired, so that all spins corresponding to $\mathrm{N_{II}}$ electrons altogether provide a zero spin. In the absence of single electrons ($\mathrm{N_{I}}=0$), the energy obviously reduces to a NOF that describes singlet states.

We focus on the mixed state of highest multiplicity: $2S+1=\mathrm{N_{I}}+1,\,S=\mathrm{N_{I}}/2$.\citep{Piris2019} For an ensemble of pure states $\left\{ \left|SM_{s}\right\rangle \right\} $, we note that the expected value of $\hat{S}_{z}$ for the whole ensemble is zero. Consequently, the spin-restricted theory can be adopted even if the total spin of the system is not zero. We use a single set of orbitals for $\alpha$ and $\beta$ spins. All the spatial orbitals will be then doubly occupied in the ensemble, so that occupancies for particles with $\alpha$ and $\beta$ spins are equal: $n_{p}^{\alpha}=n_{p}^{\beta}=n_{p}$. 

We divide the orbital space $\Omega$ into two subspaces: $\Omega=\Omega_{\mathrm{I}}\oplus\Omega_{\mathrm{II}}$. $\Omega_{\mathrm{II}}$ is composed of $\mathrm{N_{II}}/2$ mutually disjoint subspaces $\Omega{}_{g}$. Each of which contains one orbital $\left|g\right\rangle $ with $g\leq\mathrm{N_{II}}/2$, and $\mathrm{N}_{g}$ orbitals $\left|p\right\rangle $ with $p>\mathrm{N_{II}}/2$, namely,
\begin{equation} 
\Omega{}_{g}=\left\{ \left|g\right\rangle ,\left|p_{1}\right\rangle ,\left|p_{2}\right\rangle ,...,\left|p_{\mathrm{N}_{g}}\right\rangle \right\} .\label{OmegaG}
\end{equation}
Taking into account the spin, the total occupancy for a given subspace $\Omega{}_{g}$ is 2, which is reflected in the following sum rule:
\begin{equation}
\sum_{p\in\Omega_{\mathrm{II}}}n_{p}=n_{g}+\sum_{i=1}^{\mathrm{N}_{g}}n_{p_{i}}=1,\quad g=1,2,...,\frac{\mathrm{N_{II}}}{2}.\label{sum1}
\end{equation}

Here, the notation $p\in\Omega_{\mathrm{II}}$ represents all the indexes of $\left|p\right\rangle $ orbitals belonging to $\Omega_{\mathrm{II}}$. In general, $\mathrm{N}_{g}$ can be different for each subspace as long as it describes the electron pair well. For convenience, in this work we take it equal for all subspaces $\Omega{}_{g}\in\Omega_{\mathrm{II}}$ to the maximum possible value determined by the basis set used in calculations. From (\ref{sum1}), it follows that 
\begin{equation}
2\sum_{p\in\Omega_{\mathrm{II}}}n_{p}=2\sum_{g=1}^{\mathrm{N_{II}}/2}\left(n_{g}+\sum_{i=1}^{\mathrm{N}_{g}}n_{p_{i}}\right)=\mathrm{N_{II}}.\label{sumNpII}
\end{equation}
Similarly, $\Omega_{\mathrm{I}}$ is composed of $\mathrm{N_{I}}$ mutually disjoint subspaces $\Omega{}_{g}$. In contrast to $\Omega_{\mathrm{II}}$, each subspace $\Omega{}_{g}\in\Omega_{\mathrm{I}}$ contains only one orbital $g$ with $2n_{g}=1$. It is worth noting that each orbital is completely occupied individually, but we do not know whether the electron has $\alpha$ or $\beta$ spin: $n_{g}^{\alpha}=n_{g}^{\beta}=n_{g}=1/2$. It follows that 
\begin{equation}
2\sum_{p\in\Omega_{\mathrm{I}}}n_{p}=2\sum_{g=\mathrm{N_{II}}/2+1}^{\mathrm{N_{\Omega}}}n_{g}=\mathrm{N_{I}}.\label{sumNpI}
\end{equation}
In Eq. (\ref{sumNpI}), $\mathrm{\mathrm{N}_{\Omega}=}\mathrm{N_{II}}/2+\mathrm{N_{I}}$ denotes the total number of suspaces in $\Omega$. Taking into account Eqs. (\ref{sumNpII}) and (\ref{sumNpI}), the trace of the 1RDM is verified equal to the number of electrons: 
\begin{equation}
2\sum_{p\in\Omega}n_{p}=2\sum_{p\in\Omega_{\mathrm{II}}}n_{p}+2\sum_{p\in\Omega_{\mathrm{I}}}n_{p}=\mathrm{N_{II}}+\mathrm{N_{I}}=\mathrm{\mathrm{N}}.\label{norm}
\end{equation}
Using ensemble N-representability conditions, we can generate a reconstruction functional for the 2RDM in terms of the ONs that leads to GNOF: 
\begin{equation}
E=E^{intra}+E_{HF}^{inter}+E_{sta}^{inter}+E_{dyn}^{inter} \label{gnof}
\end{equation}
The intra-pair component is formed by the sum of the energies of the pairs of electrons with opposite spins and the single-electron energies of the unpaired electrons, namely 
\begin{equation}
E^{intra}=\sum\limits _{g=1}^{\mathrm{N_{II}}/2}E_{g}+{\displaystyle \sum_{g=\mathrm{N_{II}}/2+1}^{\mathrm{N}_{\Omega}}}H_{gg}
\end{equation}
\begin{equation}
\begin{array}{c}
E_{g}=\sum\limits _{p\in\Omega_{g}}n_{p}(2H_{pp}+J_{pp})\quad\\
\\
\quad\quad+\sum\limits _{q,p\in\Omega_{g},p\neq q}\Pi\left(n_{q},n_{p}\right)L_{pq}
\end{array}
\end{equation}
where 
\begin{equation}
\Pi\left(n_{q},n_{p}\right)=\sqrt{n_{q}n_{p}}\left(\delta_{q\Omega^{a}}\delta_{p\Omega^{a}}-\delta_{qg}-\delta_{pg}\right)
\end{equation}
and $H_{pp}$ are the diagonal one-electron matrix elements of the kinetic energy and external potential operators. $J_{pq}=\left\langle pq|pq\right\rangle$ and $L_{pq}=\left\langle pp|qq\right\rangle$ are the Coulomb and exchange-time-inversion integrals, respectively. $\Omega^{a}$ denotes the subspace composed of orbitals above the level $\mathrm{N}_{\Omega}$ ($p>\mathrm{N}_{\Omega}$). The inter-subspace HF term is 
\begin{equation}
E_{HF}^{inter}=\sum\limits _{p,q=1}^{\mathrm{N}_{B}}\,'\,n_{q}n_{p}\left(2J_{pq}-K_{pq}\right)\label{ehf}
\end{equation}
where $K_{pq}=\left\langle pq|qp\right\rangle $ are the exchange integrals. The prime in the summation indicates that only the inter-subspace terms are taking into account ($p\in\Omega{}_{f},q\in\Omega{}_{g},f\neq g$). $\mathrm{N}_{B}$ represents the number of basis functions considered. The inter-subspace static component is written as 
\begin{equation}
\begin{array}{c}
E_{sta}^{inter}=-\left({\displaystyle \sum_{p=1}^{\mathrm{N}_{\Omega}}\sum_{q=\mathrm{N}_{\Omega}+1}^{\mathrm{N}_{B}}+\sum_{p=\mathrm{N}_{\Omega}+1}^{\mathrm{N}_{B}}\sum_{q=1}^{\mathrm{N}_{\Omega}}}\right.\left.{\displaystyle +\sum_{p,q=\mathrm{N}_{\Omega}+1}^{\mathrm{N}_{B}}}\right)'\\
\\
\Phi_{q}\Phi_{p}L_{pq}-\:\dfrac{1}{2}\left({\displaystyle \sum\limits _{p=1}^{\mathrm{N_{II}}/2}\sum_{q=\mathrm{N_{II}}/2+1}^{\mathrm{N}_{\Omega}}+\sum_{p=\mathrm{N_{II}}/2+1}^{\mathrm{N}_{\Omega}}\sum\limits _{q=1}^{\mathrm{N_{II}}/2}}\right)'\\
\\
\Phi_{q}\Phi_{p}L_{pq}{\displaystyle \:-\:\dfrac{1}{4}\sum_{p,q=\mathrm{N_{II}}/2+1}^{\mathrm{N}_{\Omega}}}K_{pq}
\end{array} \label{esta}
\end{equation}
where $\Phi_{p}=\sqrt{n_{p}h_{p}}$ with the hole $h_{p}=1-n_{p}$. Note that $\Phi_{p}$ has significant values only when the occupation number $n_{p}$ differs substantially from 1 and 0. Finally, the inter-subspace dynamic energy can be conveniently expressed as 
\begin{equation}
\begin{array}{c}
E_{dyn}^{inter}=\sum\limits _{p,q=1}^{\mathrm{N}_{B}}\,'\,\left[n_{q}^{d}n_{p}^{d}+\;\Pi\left(n_{q}^{d},n_{p}^{d}\right)\right]\\
\\
\left(1-\delta_{q\Omega_{II}^{b}}\delta_{p\Omega_{II}^{b}}\right)L_{pq}\end{array}\label{edyn}
\end{equation}

In Eq. (\ref{edyn}), $\Omega_{II}^{b}$ denotes the subspace composed of orbitals below the level $\mathrm{N_{II}}/2$ ($p\leq\mathrm{N_{II}}/2$), so interactions between orbitals belonging to $\Omega_{II}^{b}$ are excluded from $E_{dyn}^{inter}$. The dynamic part of the ON $n_{p}$ is defined as 
\begin{equation}
n_{p}^{d}=n_{p}\cdot e^{-\left(\dfrac{h_{g}}{h_{c}}\right)^{2}},\quad p\in\Omega_{g}\ \label{dyn-on}
\end{equation}
with $h_{c}=0.02\sqrt{2}$.\citep{Piris2021b} The maximum value of $n_{p}^{d}$ is around 0.012 in accordance with the Pulay\textquoteright s criterion that establishes an occupancy deviation of approximately 0.01 with respect to 1 or 0 for a NO to contribute to the dynamic correlation. Clearly, GNOF does not take into account dynamic correlation of the single electrons ($p\in\Omega_{\mathrm{I}}$) via the $E_{dyn}^{inter}$ term. Considering real spatial orbitals ($L_{pq}=K_{pq}$) and $n_{p}\approx n_{p}^{d}$, it is not difficult to verify that the terms proportional to the product of the ONs will cancel out, so that only those terms proportional to $\Pi$ will contribute significantly to the energy.

It is important to note that GNOF preserves the total spin of the multiplet: $\expval{\hat{S}^{2}}=S\left(S+1\right)$.\citep{Piris2019}  Taking into account that GNOF does not contain inter-subspace terms between orbitals below $\mathrm{{N}_{B}}$, except for the HF-like terms of the Eq. (\ref{ehf}), Eq. (\ref{gnof}) reduces to the PNOF7-like functional\citep{Piris2017,mitxelena2018a} when the inter-pair dynamic term ($E_{dyn}^{inter}$) is neglected. Furthermore, taking $\Phi_{p}=2n_{p}h_{p}$ in Eq. (\ref{esta}) the PNOF7s-like version of the functional is obtained.\citep{Piris2018b} Finally, if the inter-subspace static term ($E_{sta}^{inter})$ is also disregarded, then GNOF reduces to PNOF5.\citep{Piris2013e}

Solutions of PNOFs can be characterized according to the recently proposed M-diagnostic\citep{Tishchenko2008-jj} adapted to the NOF multiplet calculations,\citep{Lew-Yee2022} namely,
\begin{equation}
    M = [1 - n_{LSONO}] + n_{LWONO}
    \label{eq:M-diag}
\end{equation}
where LSONO stands for the least strongly occupied NO, that is, the orbital with ON farthest from 1 below $\mathrm{N_{II}/2}$, so it belongs to $\Omega_{II}^{b}$ subspace, and LWONO for the least weakly occupied NO, that is, the orbital with ON farthest from 0 above $N_{\Omega}$, so it belongs to $\Omega^a$ subspace. Recall that M values close to zero indicate the predominance of dynamic correlation, while values beyond 0.1 indicate the predominance of static correlation.

\section{Computational Details}\label{sec:comp-details}

In this work, we have used the optimized structures of the FeP reported in Ref.~\citep{Groenhof2005-jp} for the singlet, triplet, and quintet states, as has been used in subsequent studies\citep{Lemke2022,Guo2021-cz,Lee2020-vg}, hence the energy gaps are computed adiabatically. It has been reported that the \ce{Fe-N} distance might be relevant for the energetics of the problem, in the used structures this distance correspond to 1.979~\angstrom~for the singlet, 1.976~\angstrom~for the triplet, and 2.053~\angstrom~for the quintet. The solution of the NOF equations has been established by optimizing the energy separately with respect to the ONs and to the NOs. Therefore, orbitals vary along the optimization process until the most favorable orbital interactions are found. NOF-MP2 calculations have been carried out as described in Ref.~\citep{Rodriguez-Mayorga2021}. We have taken this opportunity to test an in-house software written in Julia, currently named DoNOF.jl,\bibnote{DoNOF.jl code can be found on https://github.com/felipelewyee/DoNOF.jl} and with integral transformation accelerated by graphic processing units (GPUs) in the calculations of the perfect pairing approach ($\mathrm{N_g}=1$), while the extended pairing calculations ($\mathrm{N_g}=4$) have been carried out using the DoNOF code.\citep{Piris2021} The correlation-consistent valence double-basis set including polarization (cc-pVDZ)\citep{Dunning1989,Balabanov2005} was used throughout, as has been previously reported that the active space is more important than using a larger basis set (e.g., cc-pVTZ) to achieve the correct prediction \citep{Lemke2022,Smith2017a}. The resolution of the identity (RI) was used to reduce the computational cost of the calculations, as reported in Ref.~\citep{Lew-Yee2021-mm} (including for NOF-MP2), and the cc-pVDZ-jkfit\citep{Weigend2002-dk} auxiliary basis set was used for all atoms except iron, for which the def2-universal-jkfit\citep{Weigend2008-rp} auxiliary basis set was used. 

\section{Results and Discussion}\label{sec:results}

We aim to understand the stabilization of the spin states in terms of the static and dynamic correlation effects by means of PNOF5, PNOF7s, PNOF7, NOF-MP2, and GNOF calculations. For this purpose, a discussion is given for both the perfect pairing and the extended PNOF approaches, with special attention to the features of the solutions given by each functional.

\subsection{Perfect Pairing}

Here we study the spin-state stability of FeP using the most simple approach for electron-pairing-based NOFs, that is pairing a single weakly occupied orbital to each strongly occupied orbital in each subspace, namely the perfect-pairing approach. Table~\ref{tab:FeP-ncwo1} presents the energy values of the singlet, triplet, and quintet states of FeP in its rows, calculated with PNOF5, PNOF7s, PNOF7, NOF-MP2 and GNOF as shown in each column. First, we observe that the energy decreases according to the order PNOF5 > PNOF7s > PNOF7 > GNOF > NOF-MP2 for all spin states, which corresponds to the order of increase of electron correlation in perfect-pairing coupling. In addition, the singlet-triplet (ST) gaps and the quintet-triplet (QT) gaps allow us to check whether the spin state is more stable with respect to the triplet. Positive values indicate that the triplet state is lower in energy, whereas negative values indicate that either the singlet or the quintet state is lower in energy than the triplet state. Overall, PNOF5, PNOF7s, and PNOF7 predict the quintet as the ground state of FeP, which agrees with the multiconfigurational wavefunctions that include more static correlation, whereas, NOF-MP2 and GNOF afford the expected triplet ground state. The case of GNOF requires a more detailed analysis of the singlet state (vide infra).

\begin{table*}[ht]
\caption{\label{tab:FeP-ncwo1}Spin state energies (Hartree) for FeP calculated by a perfect pairing PNOF5, PNOF7s, PNOF7, NOF-MP2 and GNOF, with its corresponding singlet-triplet adiabatic gap (ST),  $\text{E}_\text{singlet} - \text{E}_\text{triplet}$, and quintet-triplet adiabatic gap (QT) , $\text{E}_\text{quintet} - \text{E}_\text{triplet}$, in kcal/mol. The values correspond to calculations using the optimized geometries of Ref.~\citep{Groenhof2005-jp} and the RI approximation. \bigskip{}
}
\begin{tabular}{l|ccccc}
\hline\hline
MUL &   PNOF5   &   PNOF7s  &   PNOF7   &   NOF-MP2 &   GNOF     \tabularnewline
\hline 
S   & -2245.417 & -2245.436 & -2245.989 & -2248.384 & -2247.769 \tabularnewline
T   & -2245.484 & -2245.492 & -2246.014 & -2248.456 & -2247.869 \tabularnewline
S-T &    42     &    35     &    16     &     45    &   63      \tabularnewline
\hline 
Q   & -2245.549 & -2245.560 & -2246.042 & -2248.416 & -2247.766 \tabularnewline
Q-T &   -29    &    -36     &   -17     &     25    &   65         \tabularnewline
\hline\hline 
\end{tabular}
\bigskip{}
\end{table*}

Take the PNOF5 QT gap as a reference to analyze the results obtained, and recall that it considers only static and dynamic intrapair correlation, but does not have inter-subspace correlation terms that are important for medium and large size systems. These terms are found in PNOF7s and PNOF7 leading to deeper total energy values, but predicts QT gaps with the wrong sign. It should be noted that PNOF7 predicts a lower QT gap than PNOF7s, a performance associated with the PNOF7 static overcorrelation at the equilibrium structures where the dynamic correlation predominates. In contrast, PNOF7s takes into account the correct amount of static inter-subspace correlation, therefore, its energy is in between PNOF5 and PNOF7, but the QT gap prediction is worse due to the lack of the inter-subspace dynamic correlation.

NOF-MP2 includes the dynamic correlation taking as reference the Slater determinant formed with the PNOF7s NOs,\citep{Piris2018b} and predicts the triplet as the ground state, with a QT gap of 25 kcal/mol with the expected sign. This outcome supports the thesis that dynamic correlation is crucial to predict the triplet as the ground state. In order to obtain GNOF energies, PNOF7s NOs and ONs were used as starting solutions. Since GNOF accounts for static and dynamic correlations, this functional is also capable of predicting the triplet state to be lower in energy than the quintet.  

Regarding the singlet state, we must note that all functionals provide a state with a marked multiconfigurational character as it has been reported in previous studies.\citep{Rovira1997-iz,Lee2020-vg} Remarkably, a ST gap of 17 kcal/mol is achieved by a traditional HF-MP2 calculation that is even lower than the QT gap obtained with the NOF-MP2 method. This result confirms the importance of dynamic correlation and points out the existence of a singlet with a predominant single-reference character.

It is worth noting that PNOF7s total energies shown in Table~\ref{tab:FeP-ncwo1} are well below the values obtained by Lemke et al.,\citep{Lemke2022} namely $-2244.6016$ and $-2244.6514$ for the triplet and quintet states, respectively. The latter are very close to the HF energies, and then must correspond to local minima. In contrast, our PNOF7s energies are in better agreement with the results of CAS(44,44). \citep{Smith2017a} They are also lower in energy, since they correlate 186 electrons in 184 and 182 orbitals for triplet and quintet states, respectively. Recall that in multiplet states, single-electron subspaces are made up of a single orbital with $2n_{g}=1$, while electron-paired subspaces are those that follow the perfect pairing. 

The M diagnostic of the PNOFs solutions for the spin states of FeP are shown in Table \ref{tab:1-M}. For the triplet and quintet states, PNOF5, PNOF7s and GNOF provide solutions below 0.1, which indicates that dynamic correlation is the dominant contribution. Note that PNOF5 and PNOF7s solutions are close to 0.1, indicating that the static correlation is important despite not being the dominant contribution. In contrast, the results of PNOF7 are strongly dominated by static correlation. It is noteworthy that the singlet states achieved with all functionals present a M-diagnostic value of 1.0, which in the perfect-paring approach directly indicates a di-radical character, in agreement with previous reports.\citep{Rovira1997-iz,Lee2020-vg}

\begin{table}[ht]
\caption{\label{tab:FeP-1-M} M diagnostic for the spin states of FeP computed with PNOF5, PNOF7s and PNOF7. \bigskip{}
}
\begin{tabular}{l|cccc}
\hline\hline
MUL & PNOF5& PNOF7s & PNOF7 & GNOF \tabularnewline
\hline 
S & 1.00 & 1.00 & 1.00 & 1.00 \tabularnewline
T & 0.07 & 0.07 & 0.60 & 0.04 \tabularnewline
Q & 0.07 & 0.08 & 0.56 & 0.04 \tabularnewline
\hline\hline 
\end{tabular}
\label{tab:1-M}
\end{table}

It has been stated that the NO picture can be used to earn chemical relevant information.\citep{Piris2013-oo} Following this idea, Fig.~\ref{fig:porfirine-orbitals} presents selected frontier orbitals of each NOF considered in this work for the triplet state. A gradual transformation can be observed from right (PNOF5) to left (GNOF) through an increase in correlation. The main change can be observed in the first row corresponding to the LSNO (equivalent to the HF HOMO), where the effect of the increase of electron correlation is to allow the ``d'' orbitals of the iron atom to interact with the $\pi$ orbitals of the porphyrin, as can be seen in PNOF7 and GNOF. A similar effect can be seen in the second and third rows corresponding to the single-electron NOs of the $\Omega_I$ subspace, where the ``d'' orbital of the iron atoms appears for all PNOFs, however, the NO of GNOFs spreads throughout the molecule. These results are in accordance with the results reported in Ref. \citep{LiManni2018}, where it is stated that these orbital interactions are the key factor for the correct ordering between the triplet and quintet states, as achieved by GNOF.

\begin{table*}[htbp]
    \centering
    \begin{tabular}{cccc}
    \hline
    \hline
        PNOF5 & PNOF7s & PNOF7 & GNOF\\
        \hline
        \multicolumn{4}{l}{Orbital of highest energy in the $\Omega_b$ subspace} \\ 
        \hline
        \includegraphics[scale=0.1]{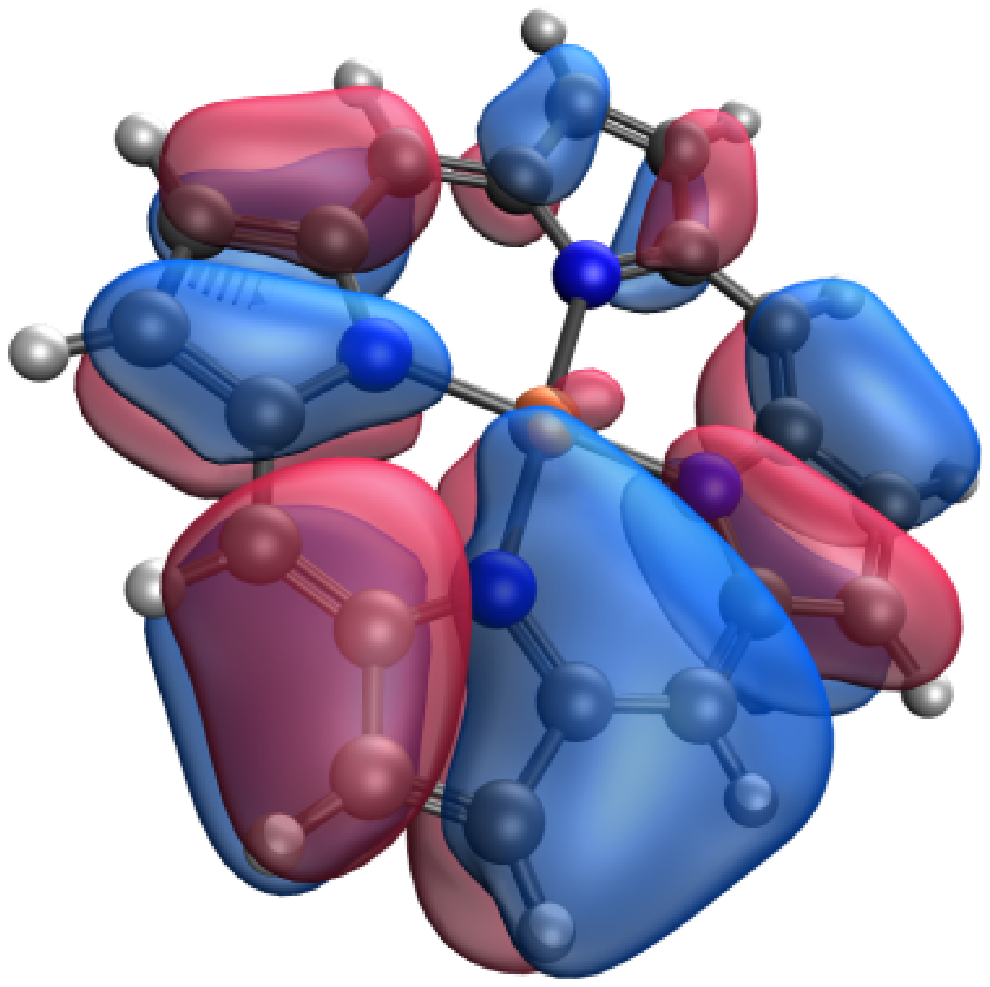} & \includegraphics[scale=0.1]{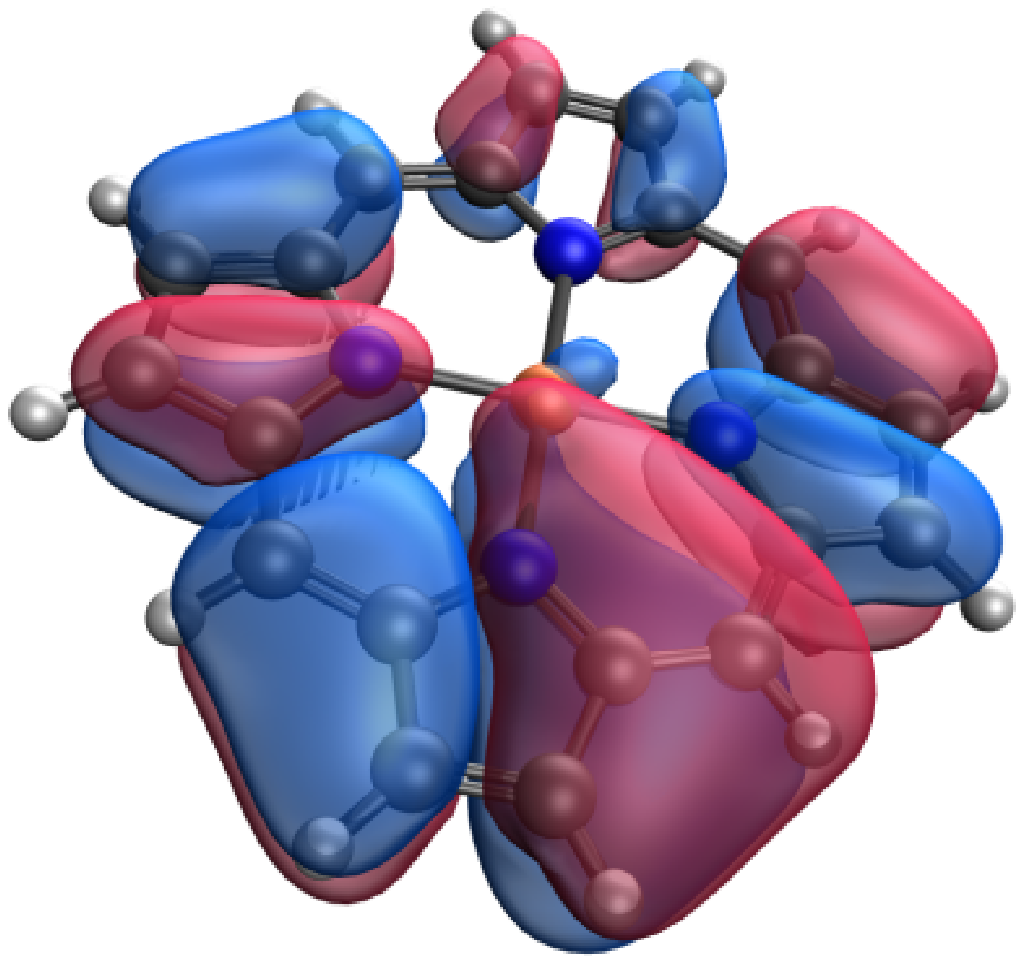} & \includegraphics[scale=0.1]{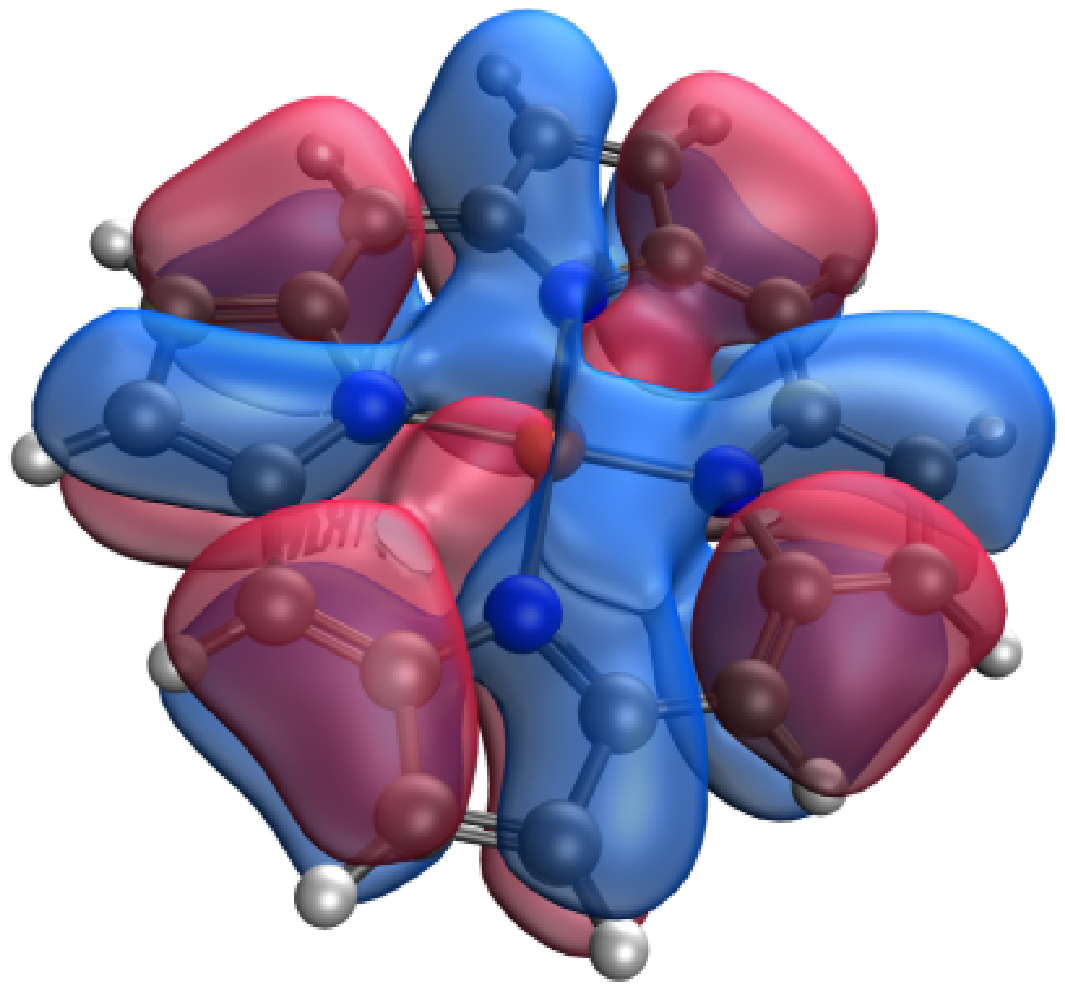} & \includegraphics[scale=0.1]{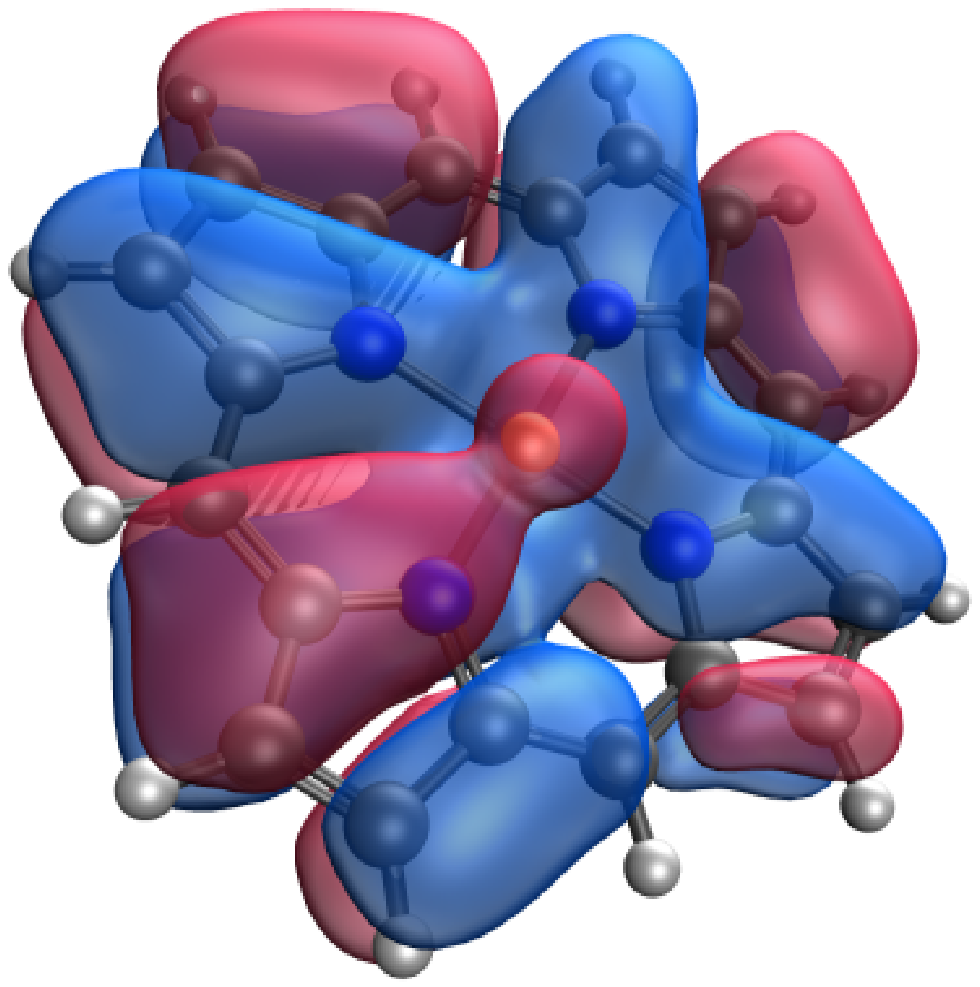}\\
        \hline
        \multicolumn{4}{l}{First orbital of the $\Omega_I$ subspace} \\
        \hline
        \includegraphics[scale=0.1]{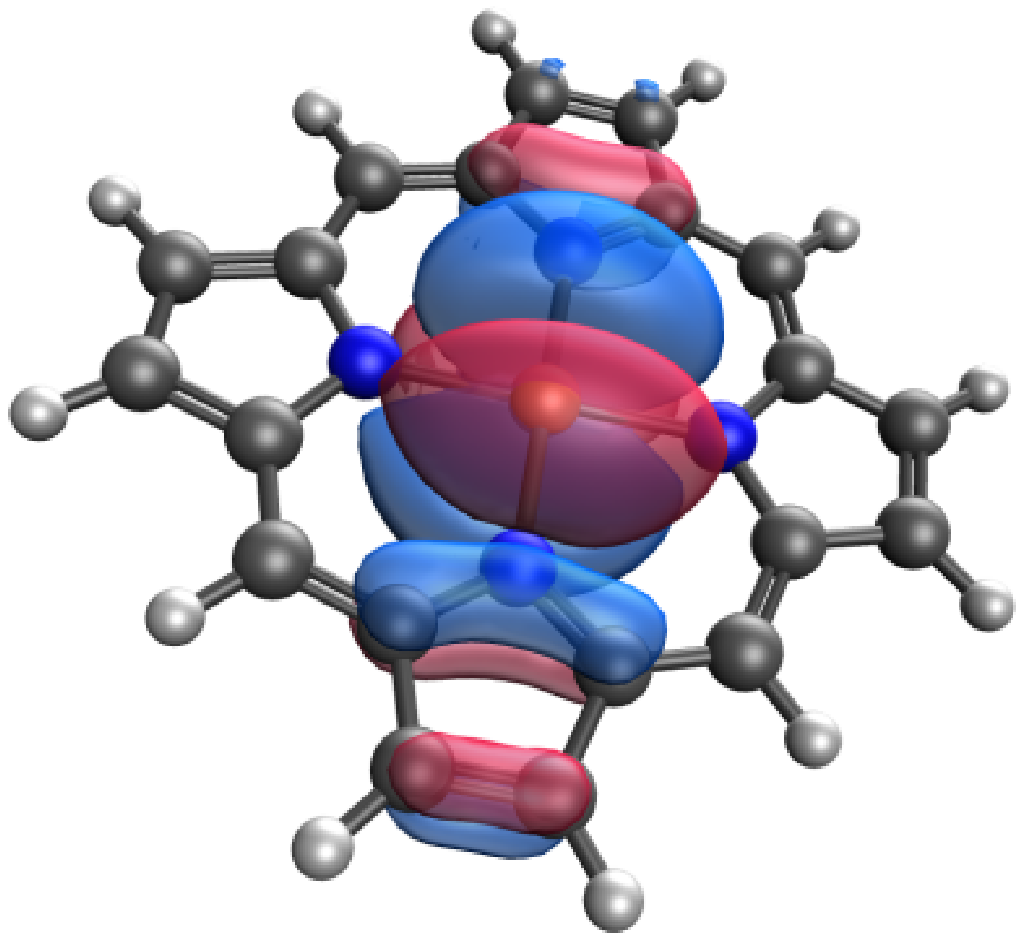} & \includegraphics[scale=0.1]{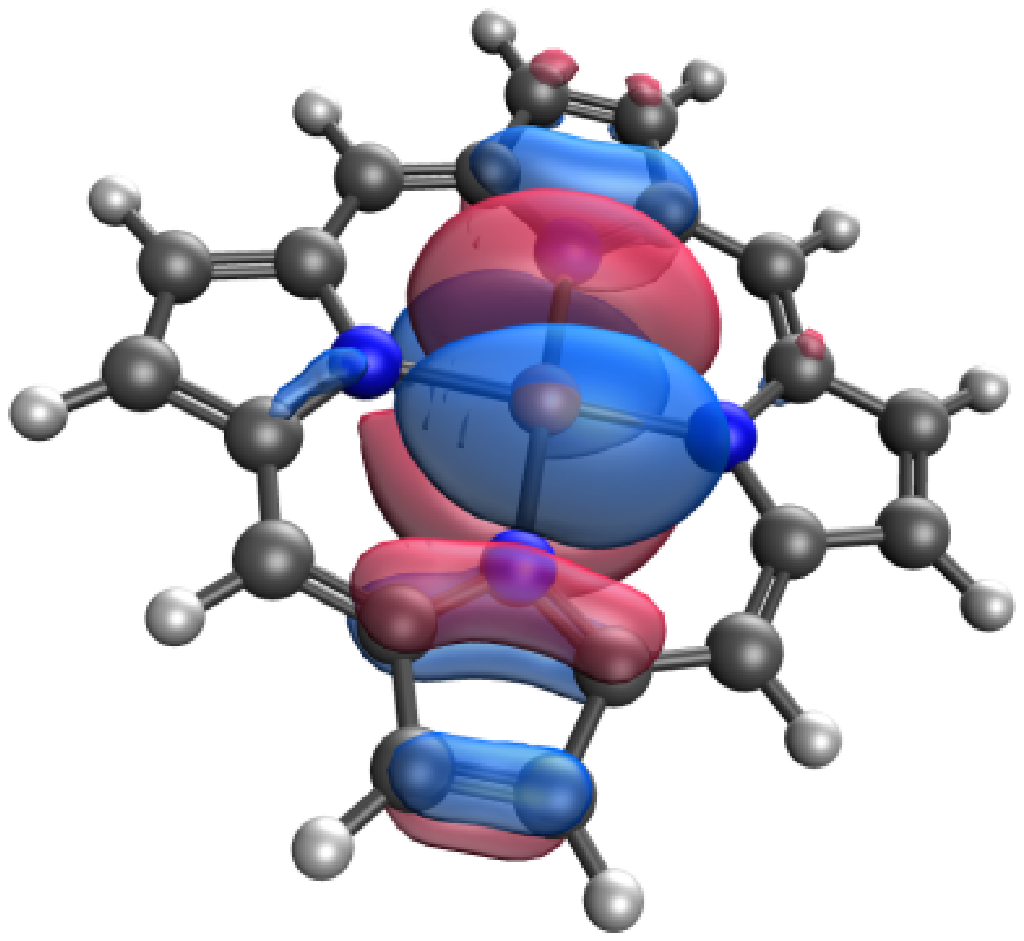} & \includegraphics[scale=0.1]{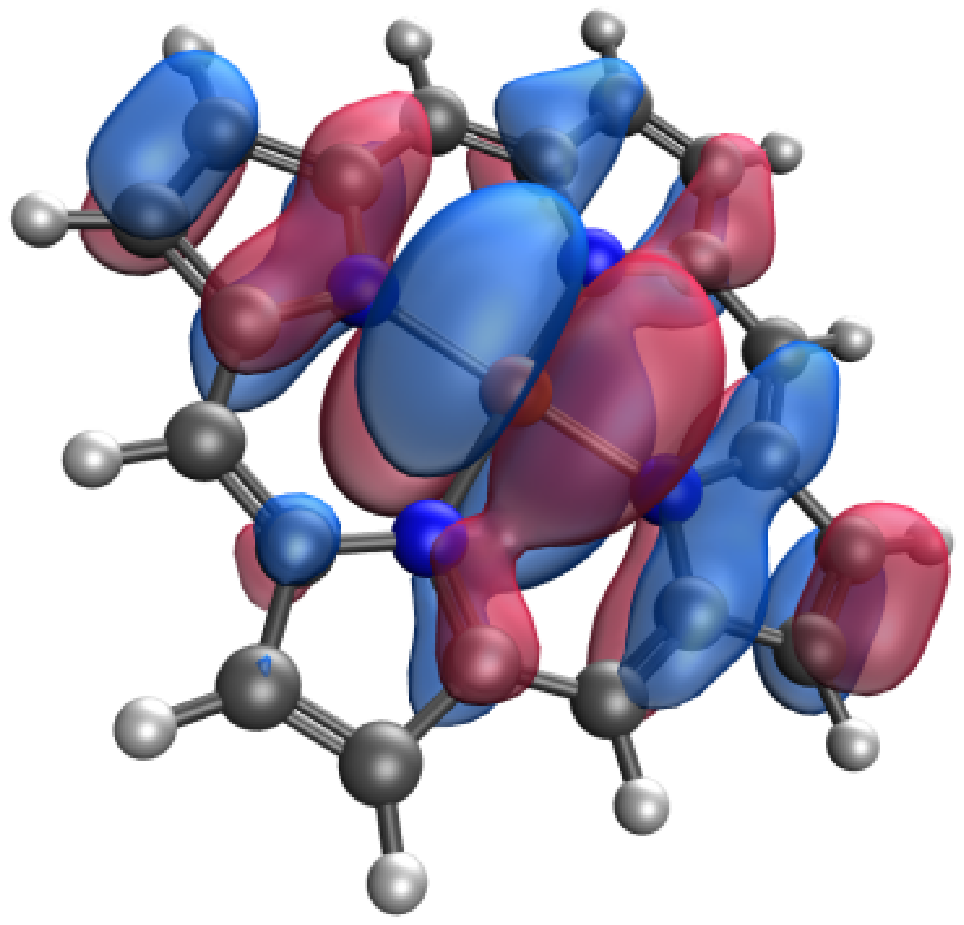} & \includegraphics[scale=0.1]{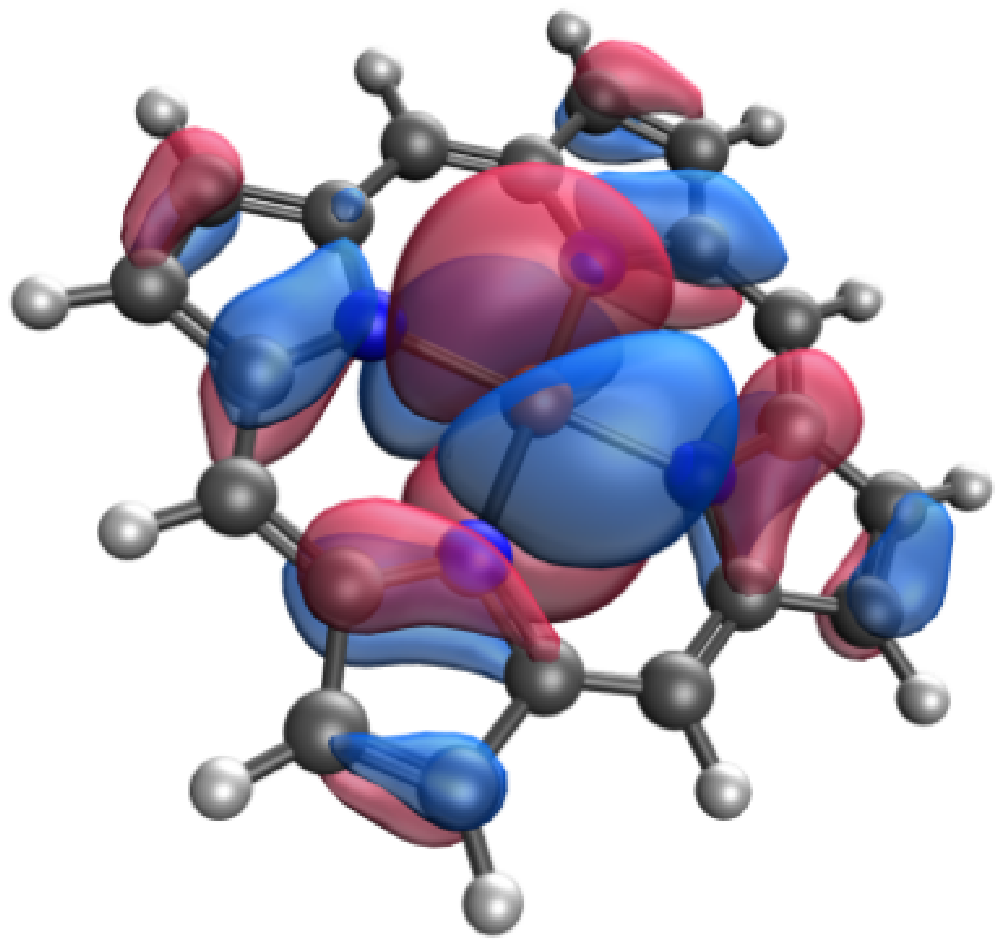}\\
        \hline
        \multicolumn{4}{l}{Second orbital of the $\Omega_I$ subspace} \\
        \hline
        \includegraphics[scale=0.1]{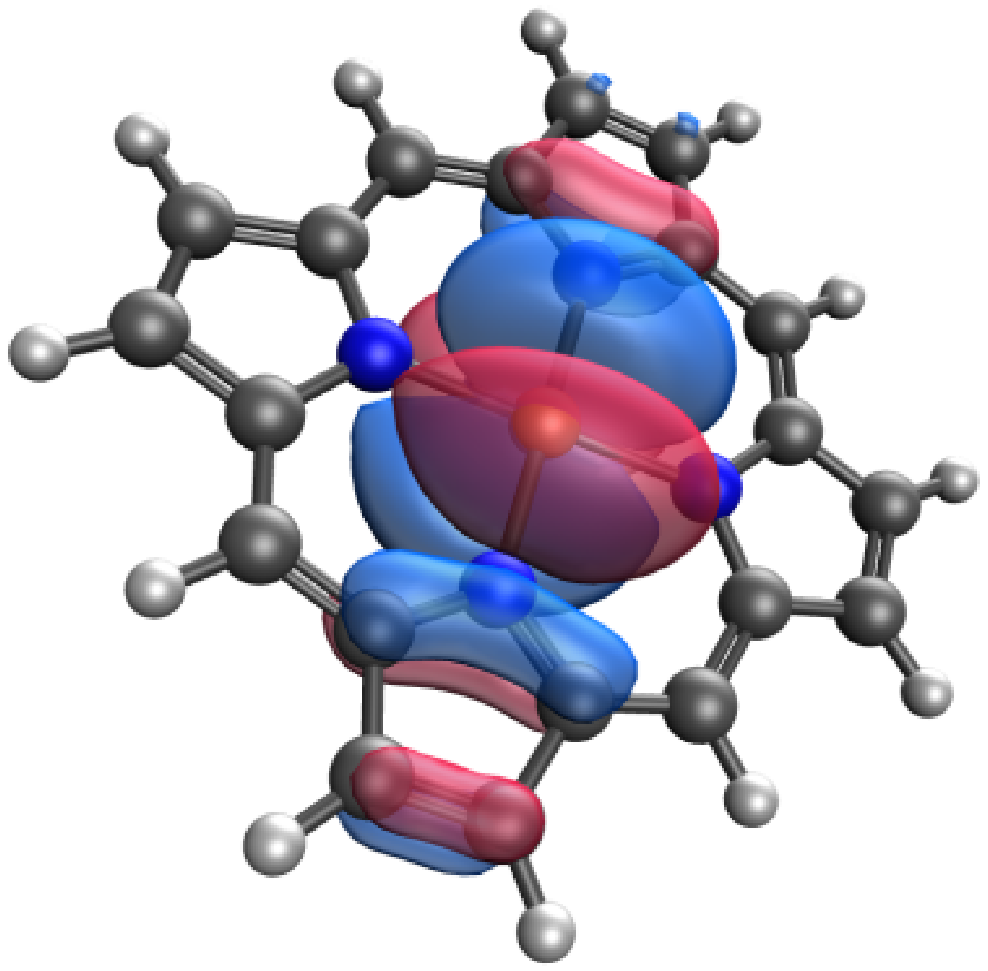} & \includegraphics[scale=0.1]{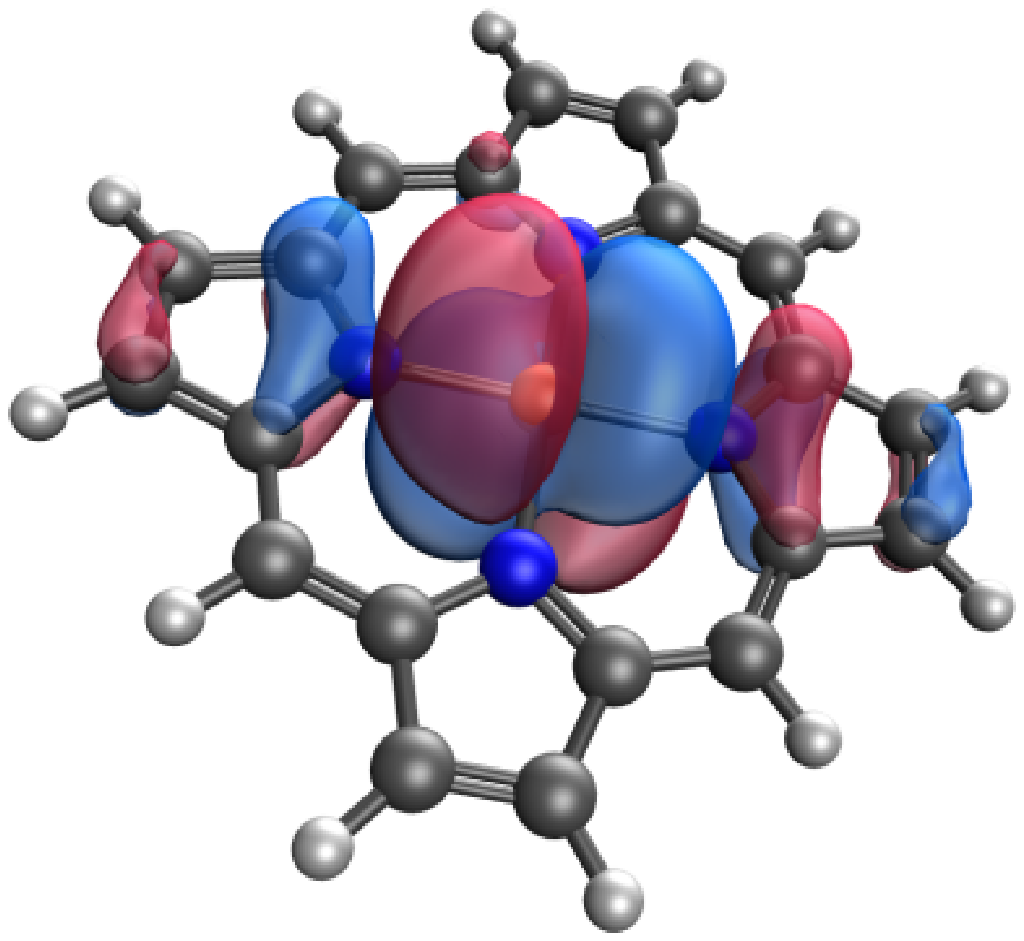} & \includegraphics[scale=0.1]{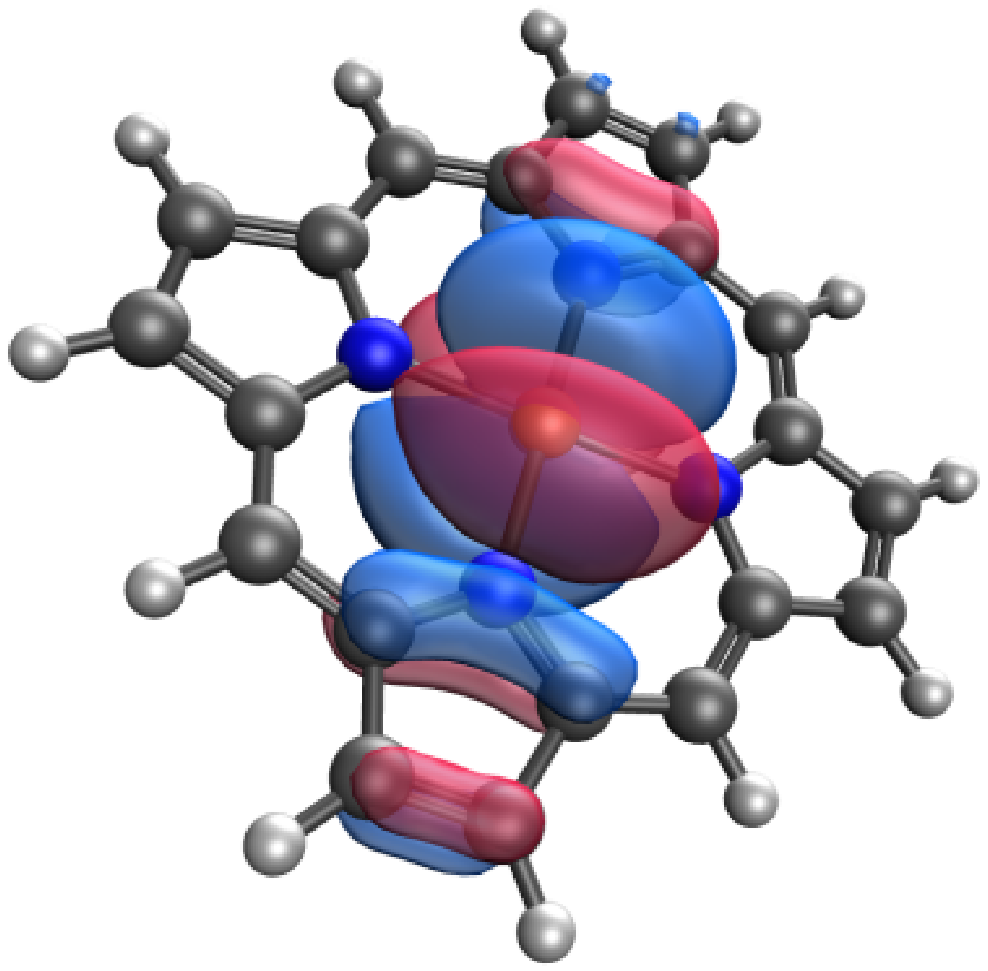} & \includegraphics[scale=0.1]{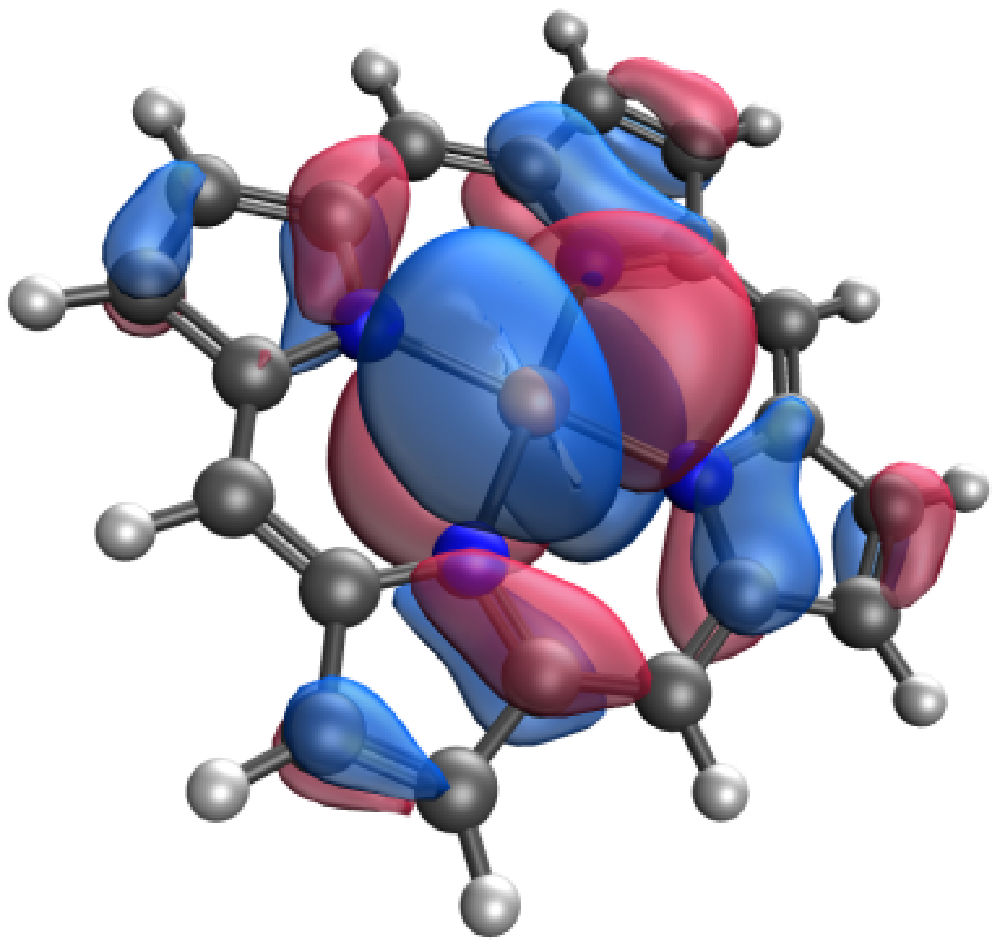}\\
        \hline
        \multicolumn{4}{l}{Orbital of lowest energy in the $\Omega_a$ subspace} \\
        \hline
        \includegraphics[scale=0.1]{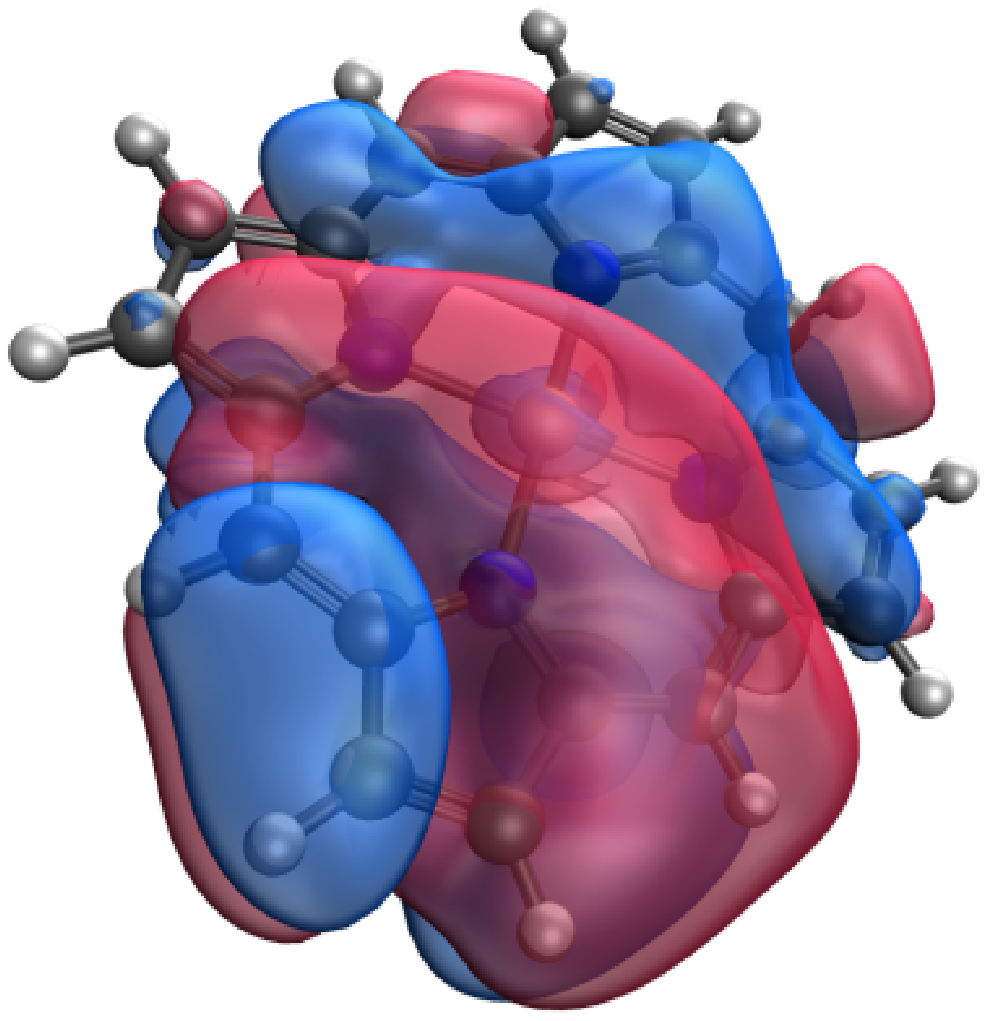} & \includegraphics[scale=0.1]{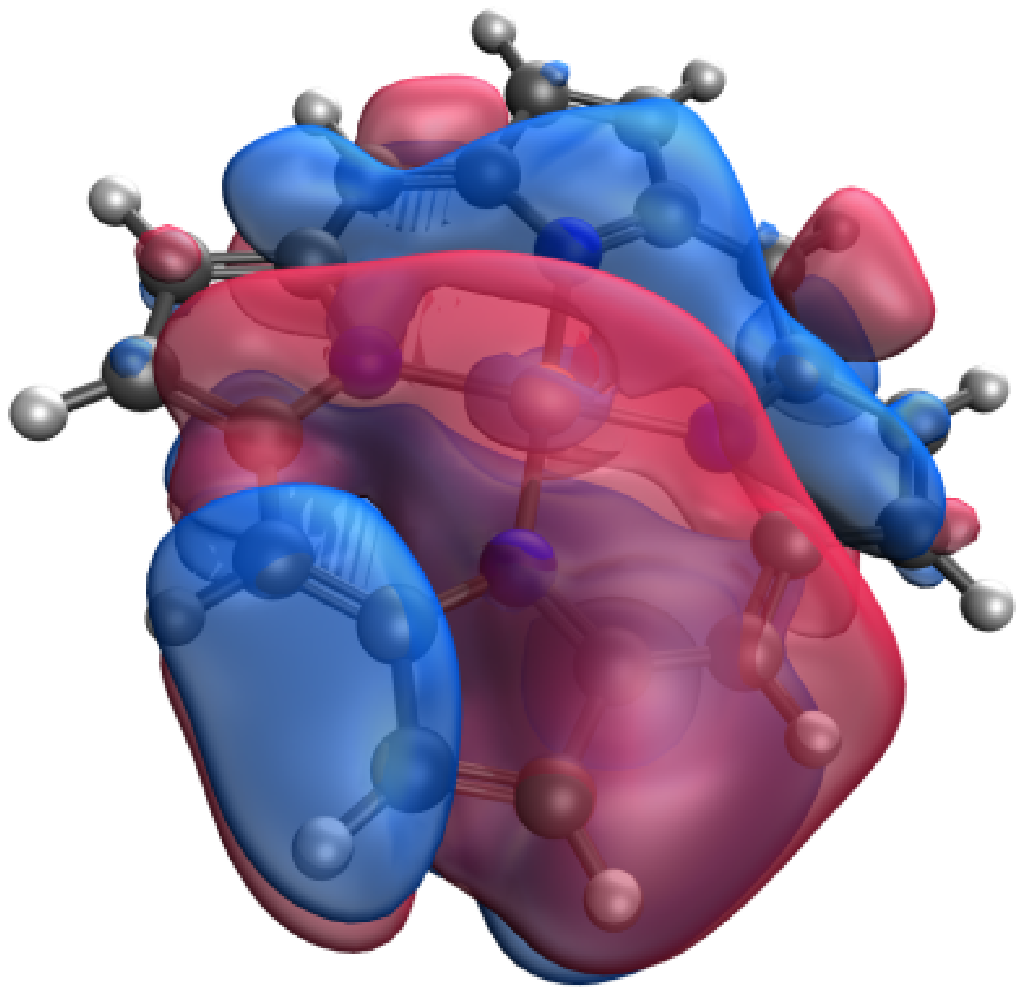} & \includegraphics[scale=0.1]{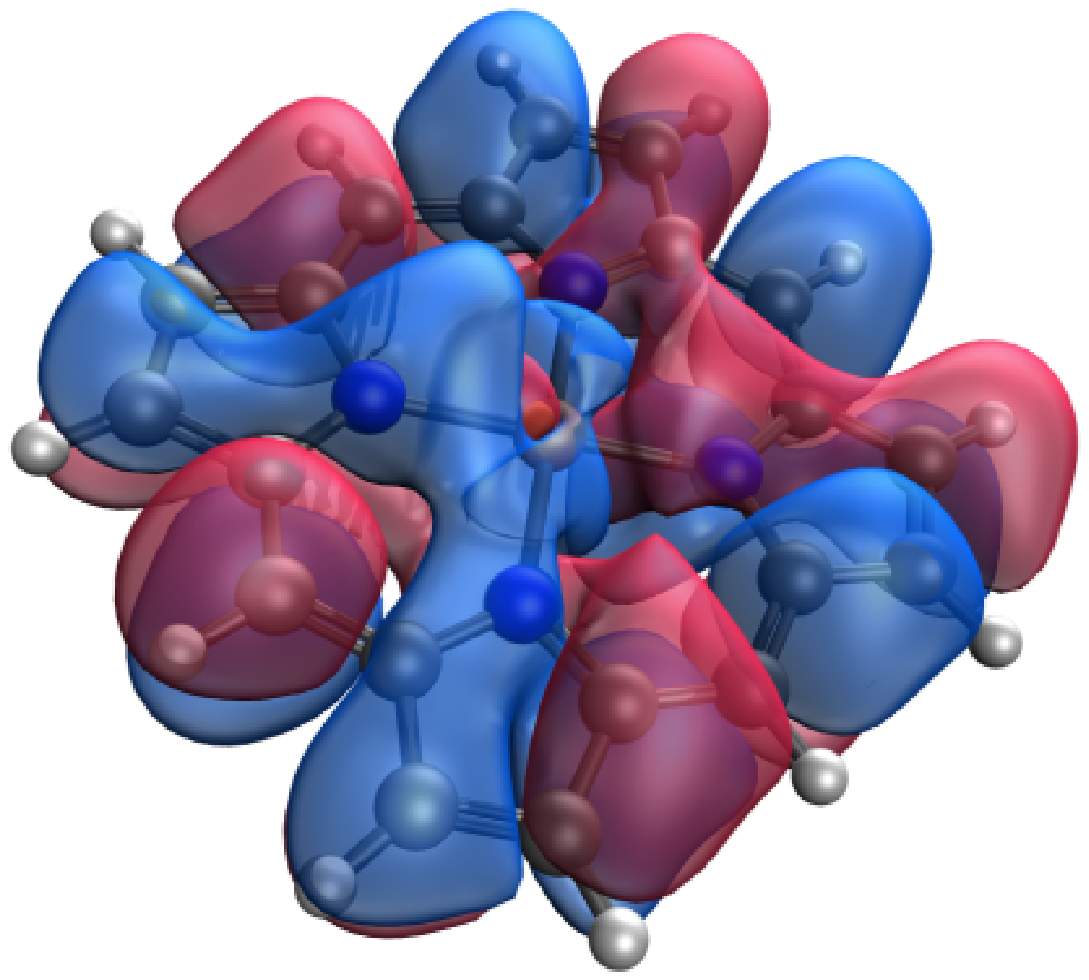} & \includegraphics[scale=0.1]{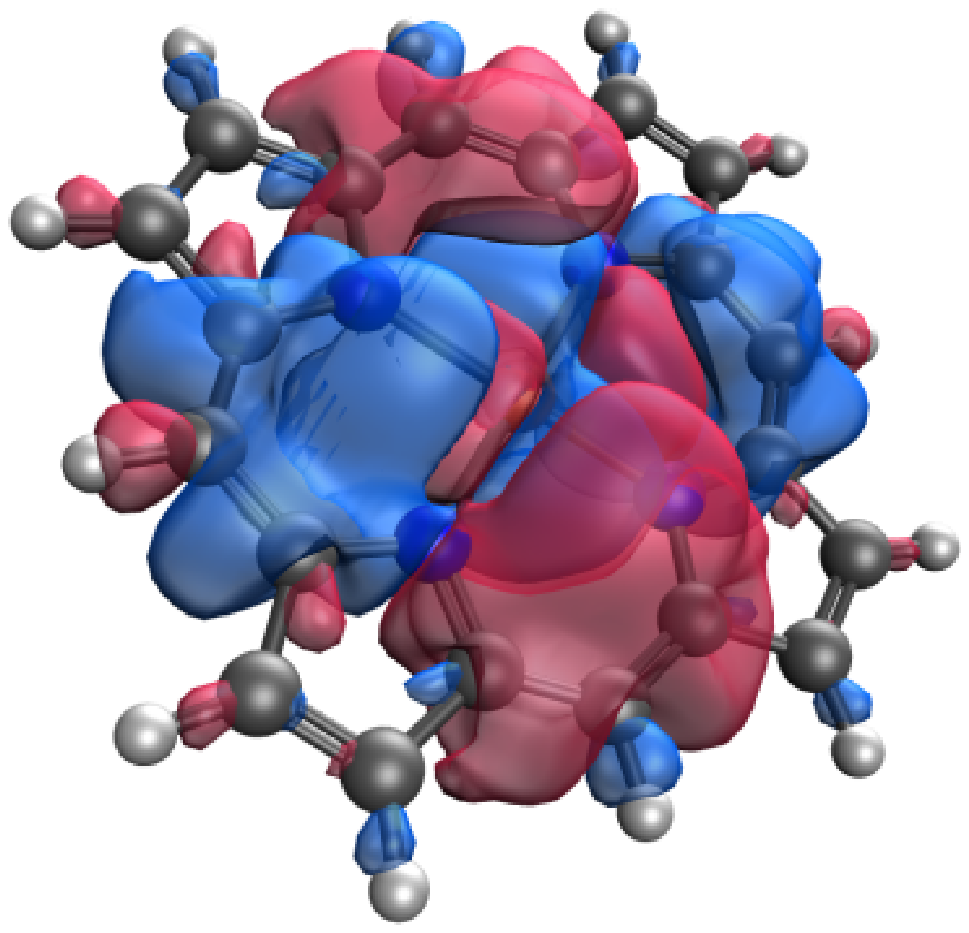}\\    
    \hline
    \hline
    \end{tabular}
    \caption{Natural orbitals of the triplet state of FeP computed with PNOF5, PNOF7s, PNOF7 and GNOF.}
    \label{fig:porfirine-orbitals}
\end{table*}

\subsection{Extended Pairing}

In this section, the extended-pairing approach is used to go beyond the results of the previous section. For this purpose, the number of weakly occupied orbitals coupled to each strongly occupied orbital was increased to four; that is, the highest possible coupling with the cc-pVDZ basis set used. Once again, we used the NOs and ONs obtained with the PNOF7s as inputs to achieve the GNOF solutions. In addition, the full electron repulsion integrals are used.

\begin{table}[H]
\caption{\label{tab:FeP-ncwo4}Spin state energies (Hartree) for FeP calculated by extended pairing PNOF5, PNOF7s, and GNOF, with its corresponding singlet-triplet adiabatic gap (ST) and quintet-triplet adiabatic gap (QT), in kcal/mol. The optimized geometries of Ref.~\citep{Groenhof2005-jp} are used.
\bigskip{}}
\begin{tabular}{l|cccc}
\hline\hline
MUL &   PNOF5   &   PNOF7s  &   GNOF     \tabularnewline
\hline 
S   & -2245.644 & -2245.696 & -2248.830   \tabularnewline
T   & -2245.742 & -2245.748 & -2248.855   \tabularnewline
S-T &    62     &    33     &    16       \tabularnewline
\hline 
Q   & -2245.766 & -2245.776 & -2248.784   \tabularnewline
Q-T &   -15     &   -16     &    45       \tabularnewline
\hline\hline 
\end{tabular}
\bigskip{}
\end{table}

It can be seen from Table~\ref{tab:FeP-ncwo4} that there is an improvement in the PNOF5 QT gap as the amount of intrapair correlation is increased. As expected, the QT gaps of PNOF7s and GNOF are significantly improved due to an increase in the electron correlation between orbitals that form the single- and paired-electron subspaces, which are not present in the independent pair approximation leading to PNOF5. As noted in the previous section, PNOF7 tends to overestimate the non-dynamic electronic correlation between subspaces in the equilibrium region, so these results were not included in the Table.

\subsubsection*{Singlet state with predominant dynamic correlation}

As the results of the previous section demonstrate, the inclusion of inter-subspace dynamic correlation is crucial for GNOF to favor the intermediate spin state over the low and high spin states. We also noted that the singlet obtained by GNOF from the PNOF7s solution has a marked multiconfigurational character. However, as we mentioned above, a traditional MP2 calculation based on the HF reference affords a ST gap that is below the NOF-MP2 result obtained from the reference multiconfigurational PNOF7s singlet. Consequently, we wonder if there is another GNOF singlet state where dynamic correlation predominates. In fact, starting from HF solutions, we have obtained GNOF singlet states with energies -2247.914 and -2248.918 Hartrees corresponding to perfect and extended couplings, respectively. Clearly, these energy values favor the singlet with predominant dynamic correlation as the lowest energy state in the GNOF case.

On the other hand, we must be cautious in claiming that this singlet is the state of minimum energy in FeP. If we look more closely at the expression (\ref{edyn}) that determines the dynamic correlation in GNOF, we can conclude that orbitals with ONs close to half do not contribute to the inter-subspace dynamic correlation; that is, it is actually a dynamic correlation term between the electron pairs. Consequently, GNOF does not contain dynamic correlation terms of the single electrons that appear in spin multiplets. This behavior has been observed in other systems, such as molecular oxygen, for which the ST gap is underestimated. In the case of FeP, it could be that GNOF is underestimating the energy of the triplet and quintet states as well as that of the open-shell singlet. It is evident that until we have an improved GNOF that includes the dynamic correlation of those orbitals with total occupancies equal to one (half in the case of the NO), we can only give a partial answer as this study shows.

\subsection{Computational Times}

This work ends with some details related to the time required for the calculations. In particular, the calculations presented in this section were performed in a Julia version of the DoNOF software.\citep{Piris2021,Lew-Yee2021} Similar codes have been reported\citep{Aroeira2022-wm} for other electronic structure methods. As noted above, the NOF equations are solved by optimizing the NOs and ONs separately, and these steps together form an outer iteration in the optimization procedure. Our implementation has been tested on an AMD Ryzen 5800 and in two GPUs, the first being a NVIDIA Turing RTX2080 and the second a NVIDIA Tesla V100. For reference, the hardware configurations are the following:

\begin{itemize}
    \item CPU-only calculations: AMD Ryzen 5800X with 8 cores-16 threads.
    \item GPU calculations:
    \begin{itemize}
        \item NVIDIA GeForce RTX 2080 with a 8 cores-16 threads AMD Ryzen 5800X CPU.
        \item NVIDIA Tesla V100 with a 4 cores Intel Xeon Gold 5122 CPU.
    \end{itemize}
\end{itemize}

The integral transformation on the CPU is currently based in Tullio.jl,\citep{Abbott2022-tp} while the transformation on the GPU depends mainly in TensorOperations.jl.\citep{Jutho2019-mp} Table~\ref{tab:times} presents the computational times for an outer iteration composed of 30 inner iterations of orbital optimization through the iterative diagonalization algorithm\citep{Piris2009a} and the ON optimization using the BFGS algorithm up to $|\text{grad}|<10^{-4}$.

\begin{table*}[htbp]
    \centering
    \begin{tabular}{c|ccc|ccc}
        \hline\hline
         Optimization Type      & \multicolumn3{c|}{Perfect Pairing ($\mathrm{N_g}=1$)} & \multicolumn3{c}{Extended Pairing ($\mathrm{N_g}=4$)} \\
                                & CPU & RTX2080 & V100 & CPU & RTX2080 & V100 \\
         
        \hline
        NO Optimization    & 253    &  10    &  8   &  762 & 22   & 42  \\
        ON Optimization &   3    &  0.2   & 0.7  &  109 & 113  & 113 \\
        \hline
        Outer Iteration      & 256    & 10.2   & 8.7  &  871 & 110  & 155 \\
        \hline\hline
    \end{tabular}
    \caption{Computational times (s) for calculations corresponding to the RI approximation using a CPU, a NVIDIA GeForce RTX2080 GPU, and a NVIDIA Tesla V100 GPU.}
    \label{tab:times}
\end{table*}

A significant improvement in computational time has been achieved in all cases with the GPU implementation, with the cuTENSOR library being a key factor for this success. The integral transformation is the dominating step in both the NO and ON minimization when the perfect pairing approach is employed, hence the time of an outer iteration is directly benefited when it is performed in a GPU, achieving a speed-up of around 25 times relative to the CPU for the GeFroce RTX2080 GPU, and 29 times for the Tesla V100 GPU. 

On the other hand, when the CPU is used for the calculations using the extended pairing approach, the NO optimization remains the bottleneck, but the contribution of the ON optimization to the time of an outer iteration increases significantly. In fact, when the GPU is introduced for the integral transformations, the NO optimization time is significantly reduced, as can be seen by going from 762 seconds on the CPU to 22 seconds in the RTX2080 hardware configuration, but the ON optimization time remains almost the same; this is the reason the speed-up is reduced to eight times for the extended pairing approach. It is worth noting that the integral transformation is performed only once in the ON optimization; hence, this is not the bottleneck but the calculation of the gradients performed in the CPU. We expect to present further details of this implementation in a future article.

\section{Conclusions}\label{sec:conclusions}

The PNOFs were used to elucidate the picture of the spin stability order of iron porphyrin. It has been found that NOFs that do not consider a significant amount of dynamic correlation, such as PNOF5, PNOF7s and PNOF7, favor the quintet as the ground state. In these functionals, the increase of the subspace size improves the results due to the inclusion of dynamic intrapair correlation, but the wrong sign of the quintet-triplet gap remains. On the other hand, methods incorporating significant amounts of dynamic correlation, such as NOF-MP2 and GNOF, achieve the correct prediction for the quintet-triplet gap of FeP, and predict the triplet as the ground state if we consider the singlet with multiconfigurational character for GNOF.

Surprisingly, there is another singlet state predicted by GNOF with a predominant dynamic correlation. In principle, this state is the one with the lowest energy, which reinforces the importance of the dynamic correlation in the stability of the iron porphyrin; however, GNOF does not contain dynamic correlation terms for the single electrons that appear in spin multiplets, so we cannot provide a definitive answer at this time, for this finding.

Larger systems, such as FeP with 37 atoms and 186 electrons, have been shown to be affordable for NOFs to handle high levels of correlation. This significantly increases the size of the systems susceptible to multiconfigurational treatment, especially when it comes to graphic processing units, such as those used in this work for the two-electron integral transformation.

In addition, GNOF correlates all electrons in all available orbitals preserving the total spin of multiplet states, which in the case of FeP using a double zeta basis set implies 186 electrons into 465 orbitals. To the best of our knowledge, such calculations have not been done so far with current wavefunction-based methods, and it is expected to become a reference calculation.

\begin{acknowledgement}
Support comes from Ministerio de Economía y Competitividad (Ref. PID2021-126714NB-I00). The authors thank for technical and human support provided by IZO-SGI SGIker of UPV/EHU and European funding (ERDF and ESF). J. F. H. Lew-Yee with CVU Grant No. 867718 acknowledges
CONACyT for the Ph.D. scholarship. J. M. del Campo acknowledges funding from projects Grant Nos. CB-2016-282791, PAPIIT-IN201822, and computing resources from the LANCAD-UNAM-DGTIC-270 project.
\end{acknowledgement}


\begin{mcitethebibliography}{103}
\providecommand*\natexlab[1]{#1}
\providecommand*\mciteSetBstSublistMode[1]{}
\providecommand*\mciteSetBstMaxWidthForm[2]{}
\providecommand*\mciteBstWouldAddEndPuncttrue
  {\def\EndOfBibitem{\unskip.}}
\providecommand*\mciteBstWouldAddEndPunctfalse
  {\let\EndOfBibitem\relax}
\providecommand*\mciteSetBstMidEndSepPunct[3]{}
\providecommand*\mciteSetBstSublistLabelBeginEnd[3]{}
\providecommand*\EndOfBibitem{}
\mciteSetBstSublistMode{f}
\mciteSetBstMaxWidthForm{subitem}{(\alph{mcitesubitemcount})}
\mciteSetBstSublistLabelBeginEnd
  {\mcitemaxwidthsubitemform\space}
  {\relax}
  {\relax}

\bibitem[Gilbert(1975)]{Gilbert1975}
Gilbert,~T.~L. {Hohenberg-Kohn theorem for nonlocal external potentials}.
  \emph{Phys. Rev. B} \textbf{1975}, \emph{12}, 2111--2120\relax
\mciteBstWouldAddEndPuncttrue
\mciteSetBstMidEndSepPunct{\mcitedefaultmidpunct}
{\mcitedefaultendpunct}{\mcitedefaultseppunct}\relax
\EndOfBibitem
\bibitem[Donnelly and Parr(1978)Donnelly, and Parr]{Donnelly1978}
Donnelly,~R.~A.; Parr,~R.~G. {Elementary properties of an energy functional of
  the first-order reduced density matrix}. \emph{J. Chem. Phys.} \textbf{1978},
  \emph{69}, 4431--4439\relax
\mciteBstWouldAddEndPuncttrue
\mciteSetBstMidEndSepPunct{\mcitedefaultmidpunct}
{\mcitedefaultendpunct}{\mcitedefaultseppunct}\relax
\EndOfBibitem
\bibitem[Levy(1979)]{Levy1979}
Levy,~M. {Universal variational functionals of electron densities, first-order
  density matrices, and natural spin-orbitals and solution of the
  v-representability problem}. \emph{Proc. Natl. Acad. Sci. USA} \textbf{1979},
  \emph{76}, 6062--6065\relax
\mciteBstWouldAddEndPuncttrue
\mciteSetBstMidEndSepPunct{\mcitedefaultmidpunct}
{\mcitedefaultendpunct}{\mcitedefaultseppunct}\relax
\EndOfBibitem
\bibitem[Valone(1980)]{Valone1980}
Valone,~S.~M. {Consequences of extending 1 matrix energy functionals pure-state
  representable to all ensemble representable 1 matrices}. \emph{J. Chem.
  Phys.} \textbf{1980}, \emph{73}, 1344--1349\relax
\mciteBstWouldAddEndPuncttrue
\mciteSetBstMidEndSepPunct{\mcitedefaultmidpunct}
{\mcitedefaultendpunct}{\mcitedefaultseppunct}\relax
\EndOfBibitem
\bibitem[Goedecker and Umrigar(2000)Goedecker, and Umrigar]{Goedecker2000}
Goedecker,~S.; Umrigar,~C.~J. In \emph{Many-electron densities Reduc. density
  matrices}; Cioslowski,~J., Ed.; Kluwer Academic/Plenum Publishers: New York,
  2000; pp 165--181\relax
\mciteBstWouldAddEndPuncttrue
\mciteSetBstMidEndSepPunct{\mcitedefaultmidpunct}
{\mcitedefaultendpunct}{\mcitedefaultseppunct}\relax
\EndOfBibitem
\bibitem[Piris(2007)]{Piris2007}
Piris,~M. In \emph{Reduced-Density-Matrix Mech. with Appl. to many-electron
  atoms Mol.}; Mazziotti,~D.~A., Ed.; John Wiley and Sons: Hoboken, New Jersey,
  USA, 2007; Vol. 134; Chapter 14, pp 387--427\relax
\mciteBstWouldAddEndPuncttrue
\mciteSetBstMidEndSepPunct{\mcitedefaultmidpunct}
{\mcitedefaultendpunct}{\mcitedefaultseppunct}\relax
\EndOfBibitem
\bibitem[Donnelly(1979)]{Donnelly1979}
Donnelly,~R.~A. {On fundamental difference between energy functionals based on
  first- and second-order density matrices}. \emph{J. Chem. Phys.}
  \textbf{1979}, \emph{71}, 2874--2879\relax
\mciteBstWouldAddEndPuncttrue
\mciteSetBstMidEndSepPunct{\mcitedefaultmidpunct}
{\mcitedefaultendpunct}{\mcitedefaultseppunct}\relax
\EndOfBibitem
\bibitem[Lude{\~{n}}a \latin{et~al.}(2013)Lude{\~{n}}a, Torres, and
  Costa]{Ludena2013}
Lude{\~{n}}a,~E.~V.; Torres,~F.~J.; Costa,~C. {Functional N-Representability in
  2-Matrix, 1-Matrix, and Density Functional Theories}. \emph{J. Mod. Phys.}
  \textbf{2013}, \emph{04}, 391--400\relax
\mciteBstWouldAddEndPuncttrue
\mciteSetBstMidEndSepPunct{\mcitedefaultmidpunct}
{\mcitedefaultendpunct}{\mcitedefaultseppunct}\relax
\EndOfBibitem
\bibitem[Piris(2018)]{Piris2018d}
Piris,~M. In \emph{Many-body approaches Differ. scales a Tribut. to N. H. March
  Occas. his 90th Birthd.}; Angilella,~G. G.~N., Amovilli,~C., Eds.; Springer:
  New York, 2018; Chapter 22, pp 261--278\relax
\mciteBstWouldAddEndPuncttrue
\mciteSetBstMidEndSepPunct{\mcitedefaultmidpunct}
{\mcitedefaultendpunct}{\mcitedefaultseppunct}\relax
\EndOfBibitem
\bibitem[Piris and Ugalde(2014)Piris, and Ugalde]{Piris2014a}
Piris,~M.; Ugalde,~J. J.~M. {Perspective on natural orbital functional theory}.
  \emph{Int. J. Quantum Chem.} \textbf{2014}, \emph{114}, 1169--1175\relax
\mciteBstWouldAddEndPuncttrue
\mciteSetBstMidEndSepPunct{\mcitedefaultmidpunct}
{\mcitedefaultendpunct}{\mcitedefaultseppunct}\relax
\EndOfBibitem
\bibitem[Pernal and Giesbertz(2016)Pernal, and Giesbertz]{Pernal2016}
Pernal,~K.; Giesbertz,~K. J.~H. {Reduced Density Matrix Functional Theory
  (RDMFT) and Linear Response Time-Dependent RDMFT (TD-RDMFT)}. \emph{Top Curr
  Chem} \textbf{2016}, \emph{368}, 125--184\relax
\mciteBstWouldAddEndPuncttrue
\mciteSetBstMidEndSepPunct{\mcitedefaultmidpunct}
{\mcitedefaultendpunct}{\mcitedefaultseppunct}\relax
\EndOfBibitem
\bibitem[Schade \latin{et~al.}(2017)Schade, Kamil, and
  Bl{\"{o}}chl]{Schade2017}
Schade,~R.; Kamil,~E.; Bl{\"{o}}chl,~P. {Reduced density-matrix functionals
  from many-particle theory}. \emph{Eur. Phys. J. Spec. Top.} \textbf{2017},
  \emph{226}, 2677--2692\relax
\mciteBstWouldAddEndPuncttrue
\mciteSetBstMidEndSepPunct{\mcitedefaultmidpunct}
{\mcitedefaultendpunct}{\mcitedefaultseppunct}\relax
\EndOfBibitem
\bibitem[Mitxelena \latin{et~al.}(2019)Mitxelena, Piris, and
  Ugalde]{MITXELENA2019}
Mitxelena,~I.; Piris,~M.; Ugalde,~J.~M. In \emph{State Art Mol. Electron.
  Struct. Comput. Correl. Methods, Basis Sets More}; Hoggan,~P., Ancarani,~U.,
  Eds.; Advances in Quantum Chemistry; Academic Press, 2019; Vol.~79; Chapter
  7, pp 155--177\relax
\mciteBstWouldAddEndPuncttrue
\mciteSetBstMidEndSepPunct{\mcitedefaultmidpunct}
{\mcitedefaultendpunct}{\mcitedefaultseppunct}\relax
\EndOfBibitem
\bibitem[Piris(2019)]{Piris2019}
Piris,~M. {Natural orbital functional for multiplets}. \emph{Phys. Rev. A}
  \textbf{2019}, \emph{100}, 32508\relax
\mciteBstWouldAddEndPuncttrue
\mciteSetBstMidEndSepPunct{\mcitedefaultmidpunct}
{\mcitedefaultendpunct}{\mcitedefaultseppunct}\relax
\EndOfBibitem
\bibitem[Benavides-Riveros and Marques(2019)Benavides-Riveros, and
  Marques]{Benavides-Riveros2019}
Benavides-Riveros,~C.~L.; Marques,~M.~A. {On the time evolution of fermionic
  occupation numbers}. \emph{J. Chem. Phys.} \textbf{2019}, \emph{151},
  044112\relax
\mciteBstWouldAddEndPuncttrue
\mciteSetBstMidEndSepPunct{\mcitedefaultmidpunct}
{\mcitedefaultendpunct}{\mcitedefaultseppunct}\relax
\EndOfBibitem
\bibitem[Cioslowski \latin{et~al.}(2019)Cioslowski, Mihalka, and
  Szabados]{Cioslowski2019}
Cioslowski,~J.; Mihalka,~Z.~E.; Szabados,~A. {Bilinear Constraints Upon the
  Correlation Contribution to the Electron-Electron Repulsion Energy as a
  Functional of the One-Electron Reduced Density Matrix}. \emph{J. Chem. Theory
  Comput.} \textbf{2019}, \emph{15}, 4862--4872\relax
\mciteBstWouldAddEndPuncttrue
\mciteSetBstMidEndSepPunct{\mcitedefaultmidpunct}
{\mcitedefaultendpunct}{\mcitedefaultseppunct}\relax
\EndOfBibitem
\bibitem[Giesbertz and Ruggenthaler(2019)Giesbertz, and
  Ruggenthaler]{Giesbertz2019}
Giesbertz,~K.~J.; Ruggenthaler,~M. {One-body reduced density-matrix functional
  theory in finite basis sets at elevated temperatures}. \emph{Phys. Rep.}
  \textbf{2019}, \emph{806}, 1--47\relax
\mciteBstWouldAddEndPuncttrue
\mciteSetBstMidEndSepPunct{\mcitedefaultmidpunct}
{\mcitedefaultendpunct}{\mcitedefaultseppunct}\relax
\EndOfBibitem
\bibitem[Gritsenko and Pernal(2019)Gritsenko, and Pernal]{Gritsenko2019}
Gritsenko,~O.~V.; Pernal,~K. {Approximating one-matrix functionals without
  generalized Pauli constraints}. \emph{Phys. Rev. A} \textbf{2019},
  \emph{100}, 012509\relax
\mciteBstWouldAddEndPuncttrue
\mciteSetBstMidEndSepPunct{\mcitedefaultmidpunct}
{\mcitedefaultendpunct}{\mcitedefaultseppunct}\relax
\EndOfBibitem
\bibitem[Lopez and Piris(2019)Lopez, and Piris]{Lopez2019}
Lopez,~X.; Piris,~M. {Performance of the NOF-MP2 method in hydrogen abstraction
  reactions}. \emph{Theor. Chem. Acc.} \textbf{2019}, \emph{138}\relax
\mciteBstWouldAddEndPuncttrue
\mciteSetBstMidEndSepPunct{\mcitedefaultmidpunct}
{\mcitedefaultendpunct}{\mcitedefaultseppunct}\relax
\EndOfBibitem
\bibitem[Quintero-Monsebaiz \latin{et~al.}(2019)Quintero-Monsebaiz, Mitxelena,
  Rodr{\'{i}}guez-Mayorga, Vela, and Piris]{Quintero-Monsebaiz2019}
Quintero-Monsebaiz,~R.; Mitxelena,~I.; Rodr{\'{i}}guez-Mayorga,~M.; Vela,~A.;
  Piris,~M. {Natural orbital functional for spin-polarized periodic systems}.
  \emph{J. Phys. Condens. Matter} \textbf{2019}, \emph{31}, 165501--8\relax
\mciteBstWouldAddEndPuncttrue
\mciteSetBstMidEndSepPunct{\mcitedefaultmidpunct}
{\mcitedefaultendpunct}{\mcitedefaultseppunct}\relax
\EndOfBibitem
\bibitem[Schilling and Schilling(2019)Schilling, and Schilling]{Schilling2019}
Schilling,~C.; Schilling,~R. {Diverging Exchange Force and Form of the Exact
  Density Matrix Functional}. \emph{Phys. Rev. Lett.} \textbf{2019},
  \emph{122}, 013001--7\relax
\mciteBstWouldAddEndPuncttrue
\mciteSetBstMidEndSepPunct{\mcitedefaultmidpunct}
{\mcitedefaultendpunct}{\mcitedefaultseppunct}\relax
\EndOfBibitem
\bibitem[Schmidt \latin{et~al.}(2019)Schmidt, Benavides-Riveros, and
  Marques]{Schmidt2019}
Schmidt,~J.; Benavides-Riveros,~C.~L.; Marques,~M. A.~L. {Reduced Density
  Matrix Functional Theory for Superconductors}. \emph{Phys. Rev. B}
  \textbf{2019}, \emph{99}, 224502\relax
\mciteBstWouldAddEndPuncttrue
\mciteSetBstMidEndSepPunct{\mcitedefaultmidpunct}
{\mcitedefaultendpunct}{\mcitedefaultseppunct}\relax
\EndOfBibitem
\bibitem[Buchholz \latin{et~al.}(2019)Buchholz, Theophilou, Nielsen,
  Ruggenthaler, and Rubio]{buchholz2019}
Buchholz,~F.; Theophilou,~I.; Nielsen,~S. E.~B.; Ruggenthaler,~M.; Rubio,~A.
  {Reduced Density-Matrix Approach to Strong Matter-Photon Interaction}.
  \emph{ACS Photonics} \textbf{2019}, \emph{6}, 2694--2711\relax
\mciteBstWouldAddEndPuncttrue
\mciteSetBstMidEndSepPunct{\mcitedefaultmidpunct}
{\mcitedefaultendpunct}{\mcitedefaultseppunct}\relax
\EndOfBibitem
\bibitem[Benavides-Riveros \latin{et~al.}(2020)Benavides-Riveros, Wolff,
  Marques, and Schilling]{Benavides-Riveros2020}
Benavides-Riveros,~C.~L.; Wolff,~J.; Marques,~M.~A.; Schilling,~C. {Reduced
  Density Matrix Functional Theory for Bosons}. \emph{Phys. Rev. Lett.}
  \textbf{2020}, \emph{124}, 180603\relax
\mciteBstWouldAddEndPuncttrue
\mciteSetBstMidEndSepPunct{\mcitedefaultmidpunct}
{\mcitedefaultendpunct}{\mcitedefaultseppunct}\relax
\EndOfBibitem
\bibitem[Giesbertz(2020)]{Giesbertz2020}
Giesbertz,~K.~J. {Implications of the unitary invariance and symmetry
  restrictions on the development of proper approximate one-body
  reduced-density-matrix functionals}. \emph{Phys. Rev. A} \textbf{2020},
  \emph{102}, 052814\relax
\mciteBstWouldAddEndPuncttrue
\mciteSetBstMidEndSepPunct{\mcitedefaultmidpunct}
{\mcitedefaultendpunct}{\mcitedefaultseppunct}\relax
\EndOfBibitem
\bibitem[Cioslowski(2020)]{Cioslowski2020}
Cioslowski,~J. {The One-Electron Reduced Density Matrix Functional Theory of
  Spin-Polarized Systems}. \emph{J. Chem. Theory Comput.} \textbf{2020},
  \emph{16}, 1578--1585\relax
\mciteBstWouldAddEndPuncttrue
\mciteSetBstMidEndSepPunct{\mcitedefaultmidpunct}
{\mcitedefaultendpunct}{\mcitedefaultseppunct}\relax
\EndOfBibitem
\bibitem[Mitxelena and Piris(2020)Mitxelena, and Piris]{Mitxelena2020}
Mitxelena,~I.; Piris,~M. {An efficient method for strongly correlated electrons
  in one dimension}. \emph{J. Phys. Condens. Matter} \textbf{2020}, \emph{32},
  17LT01\relax
\mciteBstWouldAddEndPuncttrue
\mciteSetBstMidEndSepPunct{\mcitedefaultmidpunct}
{\mcitedefaultendpunct}{\mcitedefaultseppunct}\relax
\EndOfBibitem
\bibitem[Mitxelena and Piris(2020)Mitxelena, and Piris]{Mitxelena2020b}
Mitxelena,~I.; Piris,~M. {An efficient method for strongly correlated electrons
  in two-dimensions}. \emph{J. Chem. Phys.} \textbf{2020}, \emph{152},
  064108\relax
\mciteBstWouldAddEndPuncttrue
\mciteSetBstMidEndSepPunct{\mcitedefaultmidpunct}
{\mcitedefaultendpunct}{\mcitedefaultseppunct}\relax
\EndOfBibitem
\bibitem[Mitxelena and Piris(2020)Mitxelena, and Piris]{Mitxelena2020c}
Mitxelena,~I.; Piris,~M. {Analytic gradients for spin multiplets in natural
  orbital functional theory}. \emph{J. Chem. Phys.} \textbf{2020}, \emph{153},
  044101\relax
\mciteBstWouldAddEndPuncttrue
\mciteSetBstMidEndSepPunct{\mcitedefaultmidpunct}
{\mcitedefaultendpunct}{\mcitedefaultseppunct}\relax
\EndOfBibitem
\bibitem[Lew-Yee \latin{et~al.}(2021)Lew-Yee, Piris, and del
  Campo]{Lew-Yee2021}
Lew-Yee,~J. F.~H.; Piris,~M.; del Campo,~J.~M. {Resolution of the identity
  approximation applied to PNOF correlation calculations}. \emph{J. Chem.
  Phys.} \textbf{2021}, \emph{154}, 064102\relax
\mciteBstWouldAddEndPuncttrue
\mciteSetBstMidEndSepPunct{\mcitedefaultmidpunct}
{\mcitedefaultendpunct}{\mcitedefaultseppunct}\relax
\EndOfBibitem
\bibitem[Mercero \latin{et~al.}(2021)Mercero, Ugalde, and Piris]{Mercero2021}
Mercero,~J.~M.; Ugalde,~J.~M.; Piris,~M. {Chemical reactivity studies by the
  natural orbital functional second-order M{\o}ller–Plesset (NOF-MP2) method:
  water dehydrogenation by the scandium cation}. \emph{Theor. Chem. Acc.}
  \textbf{2021}, \emph{140}, 74\relax
\mciteBstWouldAddEndPuncttrue
\mciteSetBstMidEndSepPunct{\mcitedefaultmidpunct}
{\mcitedefaultendpunct}{\mcitedefaultseppunct}\relax
\EndOfBibitem
\bibitem[Qui(2021)]{Quintero-Monsebaiz2021}
{Spectroscopic properties of open shell diatomic molecules using Piris Natural
  Orbital Functionals}. \emph{Phys. Chem. Chem. Phys.} \textbf{2021},
  \emph{19}, 2953--2963\relax
\mciteBstWouldAddEndPuncttrue
\mciteSetBstMidEndSepPunct{\mcitedefaultmidpunct}
{\mcitedefaultendpunct}{\mcitedefaultseppunct}\relax
\EndOfBibitem
\bibitem[Schilling and Pittalis(2021)Schilling, and Pittalis]{Schilling2021}
Schilling,~C.; Pittalis,~S. {Ensemble Reduced Density Matrix Functional Theory
  for Excited States and Hierarchical Generalization of Pauli's Exclusion
  Principle}. \emph{Phys. Rev. Lett.} \textbf{2021}, \emph{127}, 023001\relax
\mciteBstWouldAddEndPuncttrue
\mciteSetBstMidEndSepPunct{\mcitedefaultmidpunct}
{\mcitedefaultendpunct}{\mcitedefaultseppunct}\relax
\EndOfBibitem
\bibitem[Liebert and Schilling(2021)Liebert, and Schilling]{Liebert2021}
Liebert,~J.; Schilling,~C. {Functional theory for Bose-Einstein condensates}.
  \emph{Phys. Rev. Res.} \textbf{2021}, \emph{3}, 013282\relax
\mciteBstWouldAddEndPuncttrue
\mciteSetBstMidEndSepPunct{\mcitedefaultmidpunct}
{\mcitedefaultendpunct}{\mcitedefaultseppunct}\relax
\EndOfBibitem
\bibitem[Wang \latin{et~al.}(2021)Wang, Knowles, and Wang]{Wang2021}
Wang,~Y.; Knowles,~P.~J.; Wang,~J. {Information entropy as a measure of the
  correlation energy associated with the cumulant}. \emph{Phys. Rev. A}
  \textbf{2021}, \emph{103}, 062808\relax
\mciteBstWouldAddEndPuncttrue
\mciteSetBstMidEndSepPunct{\mcitedefaultmidpunct}
{\mcitedefaultendpunct}{\mcitedefaultseppunct}\relax
\EndOfBibitem
\bibitem[Yao \latin{et~al.}(2021)Yao, Fang, and Su]{Yao2021}
Yao,~Y.-F.; Fang,~W.-H.; Su,~N.~Q. {Handling Ensemble N-Representability
  Constraint in Explicit-by-Implicit Manner}. \emph{J. Phys. Chem. Lett.}
  \textbf{2021}, \emph{12}, 6788--6793\relax
\mciteBstWouldAddEndPuncttrue
\mciteSetBstMidEndSepPunct{\mcitedefaultmidpunct}
{\mcitedefaultendpunct}{\mcitedefaultseppunct}\relax
\EndOfBibitem
\bibitem[Gibney \latin{et~al.}(2021)Gibney, Boyn, and Mazziotti]{Gibney2021}
Gibney,~D.; Boyn,~J.~N.; Mazziotti,~D.~A. {Toward a Resolution of the Static
  Correlation Problem in Density Functional Theory from Semidefinite
  Programming}. \emph{J. Phys. Chem. Lett.} \textbf{2021}, \emph{12},
  385--391\relax
\mciteBstWouldAddEndPuncttrue
\mciteSetBstMidEndSepPunct{\mcitedefaultmidpunct}
{\mcitedefaultendpunct}{\mcitedefaultseppunct}\relax
\EndOfBibitem
\bibitem[Piris(2021)]{Piris2021b}
Piris,~M. {Global Natural Orbital Functional: Towards the Complete Description
  of the Electron Correlation}. \emph{Phys. Rev. Lett.} \textbf{2021},
  \emph{127}, 233001\relax
\mciteBstWouldAddEndPuncttrue
\mciteSetBstMidEndSepPunct{\mcitedefaultmidpunct}
{\mcitedefaultendpunct}{\mcitedefaultseppunct}\relax
\EndOfBibitem
\bibitem[{Di Sabatino} \latin{et~al.}(2022){Di Sabatino}, Koskelo, Berger, and
  Romaniello]{DiSabatino2022}
{Di Sabatino},~S.; Koskelo,~J.; Berger,~J.~A.; Romaniello,~P. {Introducing
  screening in one-body density matrix functionals: impact on the Extended
  Koopmans' Theorem's charged excitations of model systems}. \textbf{2022},
  \emph{105}, 235123\relax
\mciteBstWouldAddEndPuncttrue
\mciteSetBstMidEndSepPunct{\mcitedefaultmidpunct}
{\mcitedefaultendpunct}{\mcitedefaultseppunct}\relax
\EndOfBibitem
\bibitem[Lemke \latin{et~al.}(2022)Lemke, Kussmann, and Ochsenfeld]{Lemke2022}
Lemke,~Y.; Kussmann,~J.; Ochsenfeld,~C. {Efficient Integral-Direct Methods for
  Self-Consistent Reduced Density Matrix Functional Theory Calculations on
  Central and Graphics Processing Units}. \emph{J. Chem. Theory Comput.}
  \textbf{2022}, \emph{18}, 4229--4244\relax
\mciteBstWouldAddEndPuncttrue
\mciteSetBstMidEndSepPunct{\mcitedefaultmidpunct}
{\mcitedefaultendpunct}{\mcitedefaultseppunct}\relax
\EndOfBibitem
\bibitem[Mitxelena and Piris(2022)Mitxelena, and Piris]{Mitxelena2022}
Mitxelena,~I.; Piris,~M. {Benchmarking GNOF against FCI in challenging systems
  in one, two, and three dimensions}. \emph{J. Chem. Phys.} \textbf{2022},
  \emph{156}, 214102\relax
\mciteBstWouldAddEndPuncttrue
\mciteSetBstMidEndSepPunct{\mcitedefaultmidpunct}
{\mcitedefaultendpunct}{\mcitedefaultseppunct}\relax
\EndOfBibitem
\bibitem[Liebert \latin{et~al.}(2022)Liebert, Castillo, Labb{\'{e}}, and
  Schilling]{Liebert2022}
Liebert,~J.; Castillo,~F.; Labb{\'{e}},~J.-P.; Schilling,~C. Foundation of
  one-particle reduced density matrix functional theory for excited states.
  \emph{J. Chem. Theory Comput.} \textbf{2022}, \emph{18}, 124--140\relax
\mciteBstWouldAddEndPuncttrue
\mciteSetBstMidEndSepPunct{\mcitedefaultmidpunct}
{\mcitedefaultendpunct}{\mcitedefaultseppunct}\relax
\EndOfBibitem
\bibitem[Wang and Baerends(2022)Wang, and Baerends]{Wang2022}
Wang,~J.; Baerends,~E.~J. {Self-Consistent-Field Method for Correlated
  Many-Electron Systems with an Entropic Cumulant Energy}. \emph{Phys. Rev.
  Lett.} \textbf{2022}, \emph{128}, 013001\relax
\mciteBstWouldAddEndPuncttrue
\mciteSetBstMidEndSepPunct{\mcitedefaultmidpunct}
{\mcitedefaultendpunct}{\mcitedefaultseppunct}\relax
\EndOfBibitem
\bibitem[Ding \latin{et~al.}(2022)Ding, Liebert, and Schilling]{Ding2022}
Ding,~L.; Liebert,~J.; Schilling,~C. {Comment on "Self-Consistent-Field Method
  for Correlated Many-Electron Systems with an Entropic Cumulant Energy"}.
  \emph{arXiv 2202.05532} \textbf{2022}, \relax
\mciteBstWouldAddEndPunctfalse
\mciteSetBstMidEndSepPunct{\mcitedefaultmidpunct}
{}{\mcitedefaultseppunct}\relax
\EndOfBibitem
\bibitem[Rodr{\'{i}}guez-Mayorga \latin{et~al.}(2022)Rodr{\'{i}}guez-Mayorga,
  Giesbertz, and Visscher]{Rodriguez-Mayorga2022}
Rodr{\'{i}}guez-Mayorga,~M.; Giesbertz,~K. J.~H.; Visscher,~L. {Relativistic
  reduced density matrix functional theory}. \emph{SciPost Chem.}
  \textbf{2022}, \emph{1}, 004\relax
\mciteBstWouldAddEndPuncttrue
\mciteSetBstMidEndSepPunct{\mcitedefaultmidpunct}
{\mcitedefaultendpunct}{\mcitedefaultseppunct}\relax
\EndOfBibitem
\bibitem[Senjean \latin{et~al.}(2022)Senjean, Yalouz, Nakatani, and
  Fromager]{Senjean2022}
Senjean,~B.; Yalouz,~S.; Nakatani,~N.; Fromager,~E. {Reduced density matrix
  functional theory from an ab initio seniority-zero wave function: Exact and
  approximate formulations along an adiabatic connection path}. \emph{Phys.
  Rev. A} \textbf{2022}, \emph{106}, 032203\relax
\mciteBstWouldAddEndPuncttrue
\mciteSetBstMidEndSepPunct{\mcitedefaultmidpunct}
{\mcitedefaultendpunct}{\mcitedefaultseppunct}\relax
\EndOfBibitem
\bibitem[Lew-Yee and {M. Del Campo}(2022)Lew-Yee, and {M. Del
  Campo}]{Lew-Yee2022}
Lew-Yee,~J. F.~H.; {M. Del Campo},~J. {Charge delocalization error in Piris
  natural orbital functionals}. \emph{J. Chem. Phys.} \textbf{2022},
  \emph{157}, 104113\relax
\mciteBstWouldAddEndPuncttrue
\mciteSetBstMidEndSepPunct{\mcitedefaultmidpunct}
{\mcitedefaultendpunct}{\mcitedefaultseppunct}\relax
\EndOfBibitem
\bibitem[Piris and Mitxelena(2021)Piris, and Mitxelena]{Piris2021}
Piris,~M.; Mitxelena,~I. {DoNOF: an open-source implementation of
  natural-orbital-functional-based methods for quantum chemistry}.
  \emph{Comput. Phys. Commun.} \textbf{2021}, \emph{259}, 107651--14\relax
\mciteBstWouldAddEndPuncttrue
\mciteSetBstMidEndSepPunct{\mcitedefaultmidpunct}
{\mcitedefaultendpunct}{\mcitedefaultseppunct}\relax
\EndOfBibitem
\bibitem[Mitxelena \latin{et~al.}(2017)Mitxelena, Piris, and
  Rodriguez-Mayorga]{Mitxelena2017a}
Mitxelena,~I.; Piris,~M.; Rodriguez-Mayorga,~M. {On the performance of natural
  orbital functional approximations in the Hubbard model}. \emph{J. Phys.
  Condens. Matter} \textbf{2017}, \emph{29}, 425602\relax
\mciteBstWouldAddEndPuncttrue
\mciteSetBstMidEndSepPunct{\mcitedefaultmidpunct}
{\mcitedefaultendpunct}{\mcitedefaultseppunct}\relax
\EndOfBibitem
\bibitem[Piris(2018)]{Piris2018e}
Piris,~M. In \emph{Quantum Chem. Daw. 21st Century. Ser. Innov. Comput. Chem.};
  Carb{\'{o}}-Dorca,~R., Chakraborty,~T., Eds.; Apple Academic Press, 2018;
  Chapter 22, pp 593--620\relax
\mciteBstWouldAddEndPuncttrue
\mciteSetBstMidEndSepPunct{\mcitedefaultmidpunct}
{\mcitedefaultendpunct}{\mcitedefaultseppunct}\relax
\EndOfBibitem
\bibitem[Piris \latin{et~al.}(2011)Piris, Lopez, Ruip{\'{e}}rez, Matxain, and
  Ugalde]{Piris2011}
Piris,~M.; Lopez,~X.; Ruip{\'{e}}rez,~F.; Matxain,~J.~M.; Ugalde,~J.~M. {A
  natural orbital functional for multiconfigurational states.} \emph{J. Chem.
  Phys.} \textbf{2011}, \emph{134}, 164102\relax
\mciteBstWouldAddEndPuncttrue
\mciteSetBstMidEndSepPunct{\mcitedefaultmidpunct}
{\mcitedefaultendpunct}{\mcitedefaultseppunct}\relax
\EndOfBibitem
\bibitem[Piris \latin{et~al.}(2013)Piris, Matxain, and Lopez]{Piris2013e}
Piris,~M.; Matxain,~J.~M.; Lopez,~X. {The intrapair electron correlation in
  natural orbital functional theory}. \emph{J. Chem. Phys.} \textbf{2013},
  \emph{139}, 234109--9\relax
\mciteBstWouldAddEndPuncttrue
\mciteSetBstMidEndSepPunct{\mcitedefaultmidpunct}
{\mcitedefaultendpunct}{\mcitedefaultseppunct}\relax
\EndOfBibitem
\bibitem[Piris(2014)]{Piris2014c}
Piris,~M. {Interacting pairs in natural orbital functional theory}. \emph{J.
  Chem. Phys.} \textbf{2014}, \emph{141}, 044107\relax
\mciteBstWouldAddEndPuncttrue
\mciteSetBstMidEndSepPunct{\mcitedefaultmidpunct}
{\mcitedefaultendpunct}{\mcitedefaultseppunct}\relax
\EndOfBibitem
\bibitem[Piris(2017)]{Piris2017}
Piris,~M. {Global Method for Electron Correlation}. \emph{Phys. Rev. Lett.}
  \textbf{2017}, \emph{119}, 063002--5\relax
\mciteBstWouldAddEndPuncttrue
\mciteSetBstMidEndSepPunct{\mcitedefaultmidpunct}
{\mcitedefaultendpunct}{\mcitedefaultseppunct}\relax
\EndOfBibitem
\bibitem[Mitxelena \latin{et~al.}(2018)Mitxelena, Rodr{\'{i}}guez-Mayorga, and
  Piris]{mitxelena2018a}
Mitxelena,~I.; Rodr{\'{i}}guez-Mayorga,~M.; Piris,~M. {Phase Dilemma in Natural
  Orbital Functional Theory from the N-representability Perspective}.
  \emph{Eur. Phys. J. B} \textbf{2018}, \emph{91}, 109\relax
\mciteBstWouldAddEndPuncttrue
\mciteSetBstMidEndSepPunct{\mcitedefaultmidpunct}
{\mcitedefaultendpunct}{\mcitedefaultseppunct}\relax
\EndOfBibitem
\bibitem[Piris \latin{et~al.}(2016)Piris, Lopez, and Ugalde]{Piris2016}
Piris,~M.; Lopez,~X.; Ugalde,~J. J. M.~J. {The Bond Order of C 2 from an
  Strictly N-Representable Natural Orbital Energy Functional Perspective}.
  \emph{Chem. - A Eur. J.} \textbf{2016}, \emph{22}, 4109\relax
\mciteBstWouldAddEndPuncttrue
\mciteSetBstMidEndSepPunct{\mcitedefaultmidpunct}
{\mcitedefaultendpunct}{\mcitedefaultseppunct}\relax
\EndOfBibitem
\bibitem[Matxain \latin{et~al.}(2011)Matxain, Piris, Ruip{\'{e}}rez, Lopez, and
  Ugalde]{Matxain2011}
Matxain,~J.~M.; Piris,~M.; Ruip{\'{e}}rez,~F.; Lopez,~X.; Ugalde,~J.~M.
  {Homolytic molecular dissociation in natural orbital functional theory.}
  \emph{Phys. Chem. Chem. Phys.} \textbf{2011}, \emph{13}, 20129--20135\relax
\mciteBstWouldAddEndPuncttrue
\mciteSetBstMidEndSepPunct{\mcitedefaultmidpunct}
{\mcitedefaultendpunct}{\mcitedefaultseppunct}\relax
\EndOfBibitem
\bibitem[Ruip{\'{e}}rez \latin{et~al.}(2013)Ruip{\'{e}}rez, Piris, Ugalde, and
  Matxain]{Ruiperez2013}
Ruip{\'{e}}rez,~F.; Piris,~M.; Ugalde,~J.~M.; Matxain,~J.~M. {The natural
  orbital functional theory of the bonding in Cr(2), Mo(2) and W(2).}
  \emph{Phys. Chem. Chem. Phys.} \textbf{2013}, \emph{15}, 2055--2062\relax
\mciteBstWouldAddEndPuncttrue
\mciteSetBstMidEndSepPunct{\mcitedefaultmidpunct}
{\mcitedefaultendpunct}{\mcitedefaultseppunct}\relax
\EndOfBibitem
\bibitem[Piris(2013)]{Piris2013c}
Piris,~M. {Interpair electron correlation by second-order perturbative
  corrections to PNOF5}. \emph{J. Chem. Phys.} \textbf{2013}, \emph{139},
  064111--7\relax
\mciteBstWouldAddEndPuncttrue
\mciteSetBstMidEndSepPunct{\mcitedefaultmidpunct}
{\mcitedefaultendpunct}{\mcitedefaultseppunct}\relax
\EndOfBibitem
\bibitem[Piris \latin{et~al.}(2014)Piris, Ruip{\'{e}}rez, and
  Matxain]{Piris2014b}
Piris,~M.; Ruip{\'{e}}rez,~F.; Matxain,~J. {Assessment of the second-order
  perturbative corrections to PNOF5}. \emph{Mol. Phys.} \textbf{2014},
  \emph{112}, 711--718\relax
\mciteBstWouldAddEndPuncttrue
\mciteSetBstMidEndSepPunct{\mcitedefaultmidpunct}
{\mcitedefaultendpunct}{\mcitedefaultseppunct}\relax
\EndOfBibitem
\bibitem[Piris(2018)]{Piris2018b}
Piris,~M. {Dynamic electron-correlation energy in the
  natural-orbital-functional second-order-M{\o}ller-Plesset method from the
  orbital-invariant perturbation theory}. \emph{Phys. Rev. A} \textbf{2018},
  \emph{98}, 022504--6\relax
\mciteBstWouldAddEndPuncttrue
\mciteSetBstMidEndSepPunct{\mcitedefaultmidpunct}
{\mcitedefaultendpunct}{\mcitedefaultseppunct}\relax
\EndOfBibitem
\bibitem[Rodr{\'{i}}guez-Mayorga \latin{et~al.}(2021)Rodr{\'{i}}guez-Mayorga,
  Mitxelena, Bruneval, and Piris]{Rodriguez-Mayorga2021}
Rodr{\'{i}}guez-Mayorga,~M.; Mitxelena,~I.; Bruneval,~F.; Piris,~M. {Coupling
  Natural Orbital Functional Theory and Many-Body Perturbation Theory by Using
  Nondynamically Correlated Canonical Orbitals}. \emph{J. Chem. Theory Comput.}
  \textbf{2021}, \emph{17}, 7562--7574\relax
\mciteBstWouldAddEndPuncttrue
\mciteSetBstMidEndSepPunct{\mcitedefaultmidpunct}
{\mcitedefaultendpunct}{\mcitedefaultseppunct}\relax
\EndOfBibitem
\bibitem[Lever and Gray(1989)Lever, and Gray]{Lever1989}
Lever,~A. B.~P.; Gray,~H.~B. \emph{{Iron Porphyrins Part 3}}; Wiley-VCH: United
  Kingdom, 1989; p 322\relax
\mciteBstWouldAddEndPuncttrue
\mciteSetBstMidEndSepPunct{\mcitedefaultmidpunct}
{\mcitedefaultendpunct}{\mcitedefaultseppunct}\relax
\EndOfBibitem
\bibitem[Perutz \latin{et~al.}(1998)Perutz, Wilkinson, Paoli, and
  Dodson]{Perutz1998}
Perutz,~M.~F.; Wilkinson,~A.~J.; Paoli,~M.; Dodson,~G.~G. The stereochemical
  mechanism of the cooperative effects in hemoglobin revisited. \emph{Annu.
  ReV. Biophys. Biomol. Struct.} \textbf{1998}, \emph{27}, 1\relax
\mciteBstWouldAddEndPuncttrue
\mciteSetBstMidEndSepPunct{\mcitedefaultmidpunct}
{\mcitedefaultendpunct}{\mcitedefaultseppunct}\relax
\EndOfBibitem
\bibitem[Gouterman(1961)]{Gouterman1961-kl}
Gouterman,~M. Spectra of porphyrins. \emph{J. Mol. Spectrosc.} \textbf{1961},
  \emph{6}, 138--163\relax
\mciteBstWouldAddEndPuncttrue
\mciteSetBstMidEndSepPunct{\mcitedefaultmidpunct}
{\mcitedefaultendpunct}{\mcitedefaultseppunct}\relax
\EndOfBibitem
\bibitem[Paolesse \latin{et~al.}(2017)Paolesse, Nardis, Monti, Stefanelli, and
  Di~Natale]{Paolesse2017-un}
Paolesse,~R.; Nardis,~S.; Monti,~D.; Stefanelli,~M.; Di~Natale,~C.
  Porphyrinoids for Chemical Sensor Applications. \emph{Chem. Rev.}
  \textbf{2017}, \emph{117}, 2517--2583\relax
\mciteBstWouldAddEndPuncttrue
\mciteSetBstMidEndSepPunct{\mcitedefaultmidpunct}
{\mcitedefaultendpunct}{\mcitedefaultseppunct}\relax
\EndOfBibitem
\bibitem[Dayan and Dayan(2011)Dayan, and Dayan]{Dayan2011-zm}
Dayan,~F.; Dayan,~E. Porphyrins: One ring in the colors of life. \emph{Am.
  Sci.} \textbf{2011}, \emph{99}, 236\relax
\mciteBstWouldAddEndPuncttrue
\mciteSetBstMidEndSepPunct{\mcitedefaultmidpunct}
{\mcitedefaultendpunct}{\mcitedefaultseppunct}\relax
\EndOfBibitem
\bibitem[Obara and Kashiwagi(1982)Obara, and Kashiwagi]{Obara1982-kg}
Obara,~S.; Kashiwagi,~H. Ab initio {MO} studies of electronic states and 
  M{\"o}ssbauer spectra of high‐, intermediate‐, and low‐spin
  {Fe(II)‐porphyrin} complexes. \emph{J. Chem. Phys.} \textbf{1982},
  \emph{77}, 3155--3165\relax
\mciteBstWouldAddEndPuncttrue
\mciteSetBstMidEndSepPunct{\mcitedefaultmidpunct}
{\mcitedefaultendpunct}{\mcitedefaultseppunct}\relax
\EndOfBibitem
\bibitem[Sontum \latin{et~al.}(1983)Sontum, Case, and Karplus]{Sontum1983-bk}
Sontum,~S.~F.; Case,~D.~A.; Karplus,~M. {X$\alpha$} multiple scattering
  calculations on {iron(II}) porphine. \emph{J. Chem. Phys.} \textbf{1983},
  \emph{79}, 2881--2892\relax
\mciteBstWouldAddEndPuncttrue
\mciteSetBstMidEndSepPunct{\mcitedefaultmidpunct}
{\mcitedefaultendpunct}{\mcitedefaultseppunct}\relax
\EndOfBibitem
\bibitem[Rohmer(1985)]{Rohmer1985-uy}
Rohmer,~M.-M. Electronic ground state of {iron(II)porphyrin}. Ab initio {SCF}
  and {CI} calculations and computed electron deformation densities.
  \emph{Chem. Phys. Lett.} \textbf{1985}, \emph{116}, 44--49\relax
\mciteBstWouldAddEndPuncttrue
\mciteSetBstMidEndSepPunct{\mcitedefaultmidpunct}
{\mcitedefaultendpunct}{\mcitedefaultseppunct}\relax
\EndOfBibitem
\bibitem[Rawlings \latin{et~al.}(1985)Rawlings, Gouterman, Davidson, and
  Feller]{Rawlings1985-sk}
Rawlings,~D.~C.; Gouterman,~M.; Davidson,~E.~R.; Feller,~D. Theoretical
  investigations of the electronic states of porphyrins. {III}. Low-lying
  electronic states of {porphinatoiron(II}). \emph{Int. J. Quantum Chem.}
  \textbf{1985}, \emph{28}, 773--796\relax
\mciteBstWouldAddEndPuncttrue
\mciteSetBstMidEndSepPunct{\mcitedefaultmidpunct}
{\mcitedefaultendpunct}{\mcitedefaultseppunct}\relax
\EndOfBibitem
\bibitem[Choe \latin{et~al.}(1998)Choe, Hashimoto, Nakano, and Hirao]{Choe1998}
Choe,~Y.~K.; Hashimoto,~T.; Nakano,~H.; Hirao,~K. {Theoretical study of the
  electronic ground state of iron(II) porphine}. \emph{Chemical Physics
  Letters} \textbf{1998}, \emph{295}, 380--388\relax
\mciteBstWouldAddEndPuncttrue
\mciteSetBstMidEndSepPunct{\mcitedefaultmidpunct}
{\mcitedefaultendpunct}{\mcitedefaultseppunct}\relax
\EndOfBibitem
\bibitem[Choe \latin{et~al.}(1999)Choe, Nakajima, Hirao, and Lindh]{Choe1999}
Choe,~Y.~K.; Nakajima,~T.; Hirao,~K.; Lindh,~R. {Theoretical study of the
  electronic ground state of iron(II) porphine. II}. \emph{Journal of Chemical
  Physics} \textbf{1999}, \emph{111}, 3837--3845\relax
\mciteBstWouldAddEndPuncttrue
\mciteSetBstMidEndSepPunct{\mcitedefaultmidpunct}
{\mcitedefaultendpunct}{\mcitedefaultseppunct}\relax
\EndOfBibitem
\bibitem[Liao and Scheiner(2002)Liao, and Scheiner]{Liao2002-zy}
Liao,~M.-S.; Scheiner,~S. Electronic structure and bonding in metal porphyrins,
  metal=Fe, Co, Ni, Cu, Zn. \emph{J. Chem. Phys.} \textbf{2002}, \emph{117},
  205--219\relax
\mciteBstWouldAddEndPuncttrue
\mciteSetBstMidEndSepPunct{\mcitedefaultmidpunct}
{\mcitedefaultendpunct}{\mcitedefaultseppunct}\relax
\EndOfBibitem
\bibitem[Groenhof \latin{et~al.}(2005)Groenhof, Swart, Ehlers, and
  Lammertsma]{Groenhof2005-jp}
Groenhof,~A.~R.; Swart,~M.; Ehlers,~A.~W.; Lammertsma,~K. Electronic ground
  states of iron porphyrin and of the first species in the catalytic reaction
  cycle of cytochrome P450s. \emph{J. Phys. Chem. A} \textbf{2005}, \emph{109},
  3411--3417\relax
\mciteBstWouldAddEndPuncttrue
\mciteSetBstMidEndSepPunct{\mcitedefaultmidpunct}
{\mcitedefaultendpunct}{\mcitedefaultseppunct}\relax
\EndOfBibitem
\bibitem[Rado{\'n}(2014)]{Radon2014-th}
Rado{\'n},~M. {Spin-State} Energetics of {Heme-Related} Models from {DFT} and
  Coupled Cluster Calculations. \emph{J. Chem. Theory Comput.} \textbf{2014},
  \emph{10}, 2306--2321\relax
\mciteBstWouldAddEndPuncttrue
\mciteSetBstMidEndSepPunct{\mcitedefaultmidpunct}
{\mcitedefaultendpunct}{\mcitedefaultseppunct}\relax
\EndOfBibitem
\bibitem[Lee \latin{et~al.}(2020)Lee, Malone, and Morales]{Lee2020-vg}
Lee,~J.; Malone,~F.~D.; Morales,~M.~A. Utilizing Essential Symmetry Breaking in
  {Auxiliary-Field} Quantum Monte Carlo: Application to the Spin Gaps of the
  {C36} Fullerene and an Iron Porphyrin Model Complex. \emph{J. Chem. Theory
  Comput.} \textbf{2020}, \emph{16}, 3019--3027\relax
\mciteBstWouldAddEndPuncttrue
\mciteSetBstMidEndSepPunct{\mcitedefaultmidpunct}
{\mcitedefaultendpunct}{\mcitedefaultseppunct}\relax
\EndOfBibitem
\bibitem[Li~Manni \latin{et~al.}(2016)Li~Manni, Smart, and
  Alavi]{Li_Manni2016-jd}
Li~Manni,~G.; Smart,~S.~D.; Alavi,~A. Combining the Complete Active Space
  {Self-Consistent} Field method and the Full Configuration Interaction Quantum
  Monte Carlo within a {Super-CI} framework, with application to challenging
  metal-porphyrins. \emph{J. Chem. Theory Comput.} \textbf{2016}, \emph{12},
  1245--1258\relax
\mciteBstWouldAddEndPuncttrue
\mciteSetBstMidEndSepPunct{\mcitedefaultmidpunct}
{\mcitedefaultendpunct}{\mcitedefaultseppunct}\relax
\EndOfBibitem
\bibitem[Smith \latin{et~al.}(2017)Smith, Mussard, Holmes, and
  Sharma]{Smith2017a}
Smith,~J.~E.; Mussard,~B.; Holmes,~A.~A.; Sharma,~S. {Cheap and Near Exact
  CASSCF with Large Active Spaces}. \emph{Journal of Chemical Theory and
  Computation} \textbf{2017}, \emph{13}, 5468--5478\relax
\mciteBstWouldAddEndPuncttrue
\mciteSetBstMidEndSepPunct{\mcitedefaultmidpunct}
{\mcitedefaultendpunct}{\mcitedefaultseppunct}\relax
\EndOfBibitem
\bibitem[Pierloot \latin{et~al.}(2017)Pierloot, Phung, and
  Domingo]{Pierloot2017}
Pierloot,~K.; Phung,~Q.~M.; Domingo,~A. {Spin State Energetics in First-Row
  Transition Metal Complexes: Contribution of (3s3p) Correlation and Its
  Description by Second-Order Perturbation Theory}. \emph{Journal of Chemical
  Theory and Computation} \textbf{2017}, \emph{13}, 537--553\relax
\mciteBstWouldAddEndPuncttrue
\mciteSetBstMidEndSepPunct{\mcitedefaultmidpunct}
{\mcitedefaultendpunct}{\mcitedefaultseppunct}\relax
\EndOfBibitem
\bibitem[Weser \latin{et~al.}(2022)Weser, Guther, Ghanem, and
  Li~Manni]{Weser2022-vl}
Weser,~O.; Guther,~K.; Ghanem,~K.; Li~Manni,~G. Stochastic Generalized Active
  Space {Self-Consistent} Field: Theory and Application. \emph{J. Chem. Theory
  Comput.} \textbf{2022}, \emph{18}, 251--272\relax
\mciteBstWouldAddEndPuncttrue
\mciteSetBstMidEndSepPunct{\mcitedefaultmidpunct}
{\mcitedefaultendpunct}{\mcitedefaultseppunct}\relax
\EndOfBibitem
\bibitem[Zhou \latin{et~al.}(2019)Zhou, Gagliardi, and Truhlar]{Zhou2019-rx}
Zhou,~C.; Gagliardi,~L.; Truhlar,~D.~G. Multiconfiguration {Pair-Density}
  Functional Theory for Iron Porphyrin with {CAS}, {RAS}, and {DMRG} Active
  Spaces. \emph{J. Phys. Chem. A} \textbf{2019}, \emph{123}, 3389--3394\relax
\mciteBstWouldAddEndPuncttrue
\mciteSetBstMidEndSepPunct{\mcitedefaultmidpunct}
{\mcitedefaultendpunct}{\mcitedefaultseppunct}\relax
\EndOfBibitem
\bibitem[Antal{\'\i}k \latin{et~al.}(2020)Antal{\'\i}k, Nachtigallov{\'a}, Lo,
  Matou{\v s}ek, Lang, Legeza, Pittner, Hobza, and Veis]{Antalik2020-br}
Antal{\'\i}k,~A.; Nachtigallov{\'a},~D.; Lo,~R.; Matou{\v s}ek,~M.; Lang,~J.;
  Legeza,~{\"O}.; Pittner,~J.; Hobza,~P.; Veis,~L. Ground state of the
  Fe(ii)-porphyrin model system corresponds to quintet: a {DFT} and
  {DMRG-based} tailored {CC} study. \emph{Phys. Chem. Chem. Phys.}
  \textbf{2020}, \emph{22}, 17033--17037\relax
\mciteBstWouldAddEndPuncttrue
\mciteSetBstMidEndSepPunct{\mcitedefaultmidpunct}
{\mcitedefaultendpunct}{\mcitedefaultseppunct}\relax
\EndOfBibitem
\bibitem[Beran \latin{et~al.}(2021)Beran, Matou{\v s}ek, Hapka, Pernal, and
  Veis]{Beran2021-fe}
Beran,~P.; Matou{\v s}ek,~M.; Hapka,~M.; Pernal,~K.; Veis,~L. Density Matrix
  Renormalization Group with Dynamical Correlation via Adiabatic Connection.
  \emph{J. Chem. Theory Comput.} \textbf{2021}, \emph{17}, 7575--7585\relax
\mciteBstWouldAddEndPuncttrue
\mciteSetBstMidEndSepPunct{\mcitedefaultmidpunct}
{\mcitedefaultendpunct}{\mcitedefaultseppunct}\relax
\EndOfBibitem
\bibitem[Berryman \latin{et~al.}(2015)Berryman, Boyd, and
  Johnson]{Berryman2015-zv}
Berryman,~V. E.~J.; Boyd,~R.~J.; Johnson,~E.~R. Balancing Exchange Mixing in
  {Density-Functional} Approximations for Iron Porphyrin. \emph{J. Chem. Theory
  Comput.} \textbf{2015}, \emph{11}, 3022--3028\relax
\mciteBstWouldAddEndPuncttrue
\mciteSetBstMidEndSepPunct{\mcitedefaultmidpunct}
{\mcitedefaultendpunct}{\mcitedefaultseppunct}\relax
\EndOfBibitem
\bibitem[Swart \latin{et~al.}(2004)Swart, Groenhof, Ehlers, and
  {others}]{Swart2004-aj}
Swart,~M.; Groenhof,~A.~R.; Ehlers,~A.~W.; {others}, Validation of
  exchange-correlation functionals for spin states of iron complexes. \emph{The
  Journal of} \textbf{2004}, \emph{108}, 5479--5483\relax
\mciteBstWouldAddEndPuncttrue
\mciteSetBstMidEndSepPunct{\mcitedefaultmidpunct}
{\mcitedefaultendpunct}{\mcitedefaultseppunct}\relax
\EndOfBibitem
\bibitem[Stroscio \latin{et~al.}(2022)Stroscio, Zhou, Truhlar, and
  Gagliardi]{Stroscio2022-wu}
Stroscio,~G.~D.; Zhou,~C.; Truhlar,~D.~G.; Gagliardi,~L. Multiconfiguration
  {Pair-Density} Functional Theory Calculations of {Iron(II}) Porphyrin:
  Effects of Hybrid {Pair-Density} Functionals and Expanded {RAS} and {DMRG}
  Active Spaces on {Spin-State} Orderings. \emph{J. Phys. Chem. A}
  \textbf{2022}, \emph{126}, 3957--3963\relax
\mciteBstWouldAddEndPuncttrue
\mciteSetBstMidEndSepPunct{\mcitedefaultmidpunct}
{\mcitedefaultendpunct}{\mcitedefaultseppunct}\relax
\EndOfBibitem
\bibitem[Tishchenko \latin{et~al.}(2008)Tishchenko, Zheng, and
  Truhlar]{Tishchenko2008-jj}
Tishchenko,~O.; Zheng,~J.; Truhlar,~D.~G. Multireference model chemistries for
  thermochemical kinetics. \emph{J. Chem. Theory Comput.} \textbf{2008},
  \emph{4}, 1208--1219\relax
\mciteBstWouldAddEndPuncttrue
\mciteSetBstMidEndSepPunct{\mcitedefaultmidpunct}
{\mcitedefaultendpunct}{\mcitedefaultseppunct}\relax
\EndOfBibitem
\bibitem[Guo \latin{et~al.}(2021)Guo, Zhang, Lei, and Liu]{Guo2021-cz}
Guo,~Y.; Zhang,~N.; Lei,~Y.; Liu,~W. {iCISCF}: An Iterative Configuration
  {Interaction-Based} Multiconfigurational {Self-Consistent} Field Theory for
  Large Active Spaces. \emph{J. Chem. Theory Comput.} \textbf{2021}, \emph{17},
  7545--7561\relax
\mciteBstWouldAddEndPuncttrue
\mciteSetBstMidEndSepPunct{\mcitedefaultmidpunct}
{\mcitedefaultendpunct}{\mcitedefaultseppunct}\relax
\EndOfBibitem
\bibitem[Not()]{Note-1}
DoNOF.jl code can be found on https://github.com/felipelewyee/DoNOF.jl\relax
\mciteBstWouldAddEndPuncttrue
\mciteSetBstMidEndSepPunct{\mcitedefaultmidpunct}
{\mcitedefaultendpunct}{\mcitedefaultseppunct}\relax
\EndOfBibitem
\bibitem[Dunning and {Dunning Jr.}(1989)Dunning, and {Dunning
  Jr.}]{Dunning1989}
Dunning,~T.~H.; {Dunning Jr.},~T.~H. {Gaussian basis sets for use in correlated
  molecular calculations. I. The atoms boron through neon and hydrogen}.
  \emph{J. Chem. Phys.} \textbf{1989}, \emph{90}, 1007--1023\relax
\mciteBstWouldAddEndPuncttrue
\mciteSetBstMidEndSepPunct{\mcitedefaultmidpunct}
{\mcitedefaultendpunct}{\mcitedefaultseppunct}\relax
\EndOfBibitem
\bibitem[Balabanov and Peterson(2005)Balabanov, and Peterson]{Balabanov2005}
Balabanov,~N.~B.; Peterson,~K.~A. {Systematically convergent basis sets for
  transition metals. I. All-electron correlation consistent basis sets for the
  3d elements Sc-Zn}. \emph{J. Chem. Phys.} \textbf{2005}, \emph{123},
  064107--15\relax
\mciteBstWouldAddEndPuncttrue
\mciteSetBstMidEndSepPunct{\mcitedefaultmidpunct}
{\mcitedefaultendpunct}{\mcitedefaultseppunct}\relax
\EndOfBibitem
\bibitem[Lew-Yee \latin{et~al.}(2021)Lew-Yee, Piris, and
  M~Del~Campo]{Lew-Yee2021-mm}
Lew-Yee,~J. F.~H.; Piris,~M.; M~Del~Campo,~J. Resolution of the identity
  approximation applied to {PNOF} correlation calculations. \emph{J. Chem.
  Phys.} \textbf{2021}, \emph{154}, 064102\relax
\mciteBstWouldAddEndPuncttrue
\mciteSetBstMidEndSepPunct{\mcitedefaultmidpunct}
{\mcitedefaultendpunct}{\mcitedefaultseppunct}\relax
\EndOfBibitem
\bibitem[Weigend(2002)]{Weigend2002-dk}
Weigend,~F. A fully direct {RI-HF} algorithm: Implementation, optimised
  auxiliary basis sets, demonstration of accuracy and efficiency. \emph{Phys.
  Chem. Chem. Phys.} \textbf{2002}, \emph{4}, 4285--4291\relax
\mciteBstWouldAddEndPuncttrue
\mciteSetBstMidEndSepPunct{\mcitedefaultmidpunct}
{\mcitedefaultendpunct}{\mcitedefaultseppunct}\relax
\EndOfBibitem
\bibitem[Weigend(2008)]{Weigend2008-rp}
Weigend,~F. {Hartree-Fock} exchange fitting basis sets for {H} to Rn. \emph{J.
  Comput. Chem.} \textbf{2008}, \emph{29}, 167--175\relax
\mciteBstWouldAddEndPuncttrue
\mciteSetBstMidEndSepPunct{\mcitedefaultmidpunct}
{\mcitedefaultendpunct}{\mcitedefaultseppunct}\relax
\EndOfBibitem
\bibitem[Rovira \latin{et~al.}(1997)Rovira, Kunc, Hutter, Ballone, and
  Parrinello]{Rovira1997-iz}
Rovira,~C.; Kunc,~K.; Hutter,~J.; Ballone,~P.; Parrinello,~M. Equilibrium
  Geometries and Electronic Structure of {Iron-Porphyrin} Complexes: A Density
  Functional Study. \emph{J. Phys. Chem. A} \textbf{1997}, \emph{101},
  8914--8925\relax
\mciteBstWouldAddEndPuncttrue
\mciteSetBstMidEndSepPunct{\mcitedefaultmidpunct}
{\mcitedefaultendpunct}{\mcitedefaultseppunct}\relax
\EndOfBibitem
\bibitem[Piris \latin{et~al.}(2013)Piris, Matxain, Lopez, and
  Ugalde]{Piris2013-oo}
Piris,~M.; Matxain,~J.~M.; Lopez,~X.; Ugalde,~J.~M. The one-electron picture in
  the Piris natural orbital functional 5 ({PNOF5}). \emph{Theor. Chem. Acc.}
  \textbf{2013}, \emph{132}, 1298\relax
\mciteBstWouldAddEndPuncttrue
\mciteSetBstMidEndSepPunct{\mcitedefaultmidpunct}
{\mcitedefaultendpunct}{\mcitedefaultseppunct}\relax
\EndOfBibitem
\bibitem[Manni and Alavi(2018)Manni, and Alavi]{LiManni2018}
Manni,~G.~L.; Alavi,~A. {Understanding the Mechanism Stabilizing Intermediate
  Spin States in Fe(II)-Porphyrin}. \emph{Journal of Physical Chemistry A}
  \textbf{2018}, \emph{122}, 4935--4947\relax
\mciteBstWouldAddEndPuncttrue
\mciteSetBstMidEndSepPunct{\mcitedefaultmidpunct}
{\mcitedefaultendpunct}{\mcitedefaultseppunct}\relax
\EndOfBibitem
\bibitem[Aroeira \latin{et~al.}(2022)Aroeira, Davis, Turney, and
  Schaefer]{Aroeira2022-wm}
Aroeira,~G. J.~R.; Davis,~M.~M.; Turney,~J.~M.; Schaefer,~H.~F.,~3rd Fermi.jl:
  A Modern Design for Quantum Chemistry. \emph{J. Chem. Theory Comput.}
  \textbf{2022}, \emph{18}, 677--686\relax
\mciteBstWouldAddEndPuncttrue
\mciteSetBstMidEndSepPunct{\mcitedefaultmidpunct}
{\mcitedefaultendpunct}{\mcitedefaultseppunct}\relax
\EndOfBibitem
\bibitem[Abbott \latin{et~al.}(2022)Abbott, Aluthge, {N3N}, Schaub, Elrod,
  Lucibello, and Chen]{Abbott2022-tp}
Abbott,~M.; Aluthge,~D.; {N3N},; Schaub,~S.; Elrod,~C.; Lucibello,~C.; Chen,~J.
  mcabbott/Tullio.jl: v0.3.5. 2022;
  \url{https://github.com/mcabbott/Tullio.jl}\relax
\mciteBstWouldAddEndPuncttrue
\mciteSetBstMidEndSepPunct{\mcitedefaultmidpunct}
{\mcitedefaultendpunct}{\mcitedefaultseppunct}\relax
\EndOfBibitem
\bibitem[{Jutho} \latin{et~al.}(2019){Jutho}, {getzdan}, Lyon, Protter, Marcus,
  {Leo}, Garrison, Otto, Saba, Iouchtchenko, Privett, and Morley]{Jutho2019-mp}
{Jutho},; {getzdan},; Lyon,~S.; Protter,~M.; Marcus,~P.~S.; {Leo},;
  Garrison,~J.; Otto,~F.; Saba,~E.; Iouchtchenko,~D.; Privett,~A.; Morley,~A.
  {TensorOperations.jl}: v1.1.0. 2019;
  \url{https://github.com/Jutho/TensorOperations.jl}\relax
\mciteBstWouldAddEndPuncttrue
\mciteSetBstMidEndSepPunct{\mcitedefaultmidpunct}
{\mcitedefaultendpunct}{\mcitedefaultseppunct}\relax
\EndOfBibitem
\bibitem[Piris and Ugalde(2009)Piris, and Ugalde]{Piris2009a}
Piris,~M.; Ugalde,~J.~M. {Iterative Diagonalization for Orbital Optimization in
  Natural Orbital Functional Theory}. \emph{J. Comput. Chem.} \textbf{2009},
  \emph{30}, 2078--2086\relax
\mciteBstWouldAddEndPuncttrue
\mciteSetBstMidEndSepPunct{\mcitedefaultmidpunct}
{\mcitedefaultendpunct}{\mcitedefaultseppunct}\relax
\EndOfBibitem
\end{mcitethebibliography}

\providecommand{\latin}[1]{#1}
\makeatletter
\providecommand{\doi}
  {\begingroup\let\do\@makeother\dospecials
  \catcode`\{=1 \catcode`\}=2 \doi@aux}
\providecommand{\doi@aux}[1]{\endgroup\texttt{#1}}
\makeatother
\providecommand*\mcitethebibliography{\thebibliography}
\csname @ifundefined\endcsname{endmcitethebibliography}
  {\let\endmcitethebibliography\endthebibliography}{}

\end{document}